%% file: binWidthCorr.tex
\RequirePackage{lineno}

\documentclass[aps,prc,showpacs,preprintnumbers,amsmath,amssymb,superscriptaddress]{revtex4-1}

\pdfoutput=1

\usepackage{graphicx}				
\usepackage{amsmath}
\usepackage{amssymb}
\usepackage{dcolumn}
\usepackage{bm}
\usepackage{color}

\graphicspath{
 {IMAGES/}
}

\input{defs}
\begin{document}
\title{Effect of centrality bin width corrections on two-particle number and transverse momentum differential correlation functions }
\author{Victor Gonzalez}
\email{victor.gonzalez@ucm.es}
\affiliation{Universidad Complutense de Madrid, Madrid 28040, Spain}
\affiliation{GSI Helmholtzzentrum f\"ur Schwerionenforschung GmbH,
Research Division and Extreme Matter Institute EMMI, 64291 Darmstadt, Germany}
\author{Ana Marin}
\affiliation{GSI Helmholtzzentrum f\"ur Schwerionenforschung,
Research Division and ExtreMe Matter Institute EMMI, Darmstadt, Germany}
\author{Pedro Ladron de Guevara}
\affiliation{Universidad Complutense de Madrid, Spain}
\affiliation{Instituto de F\'{\i}sica, Universidad Nacional Aut\'{o}noma de M\'{e}xico,
CP 04510, CDMX, Mexico}

\author{Jinjin Pan}
\email{jinjin.pan@cern.ch}
\affiliation{Department of Physics and Astronomy, Wayne State University, Detroit, Michigan 48201, USA}
\author{Sumit Basu}
\email{sumits.basu@cern.ch}
\affiliation{Department of Physics and Astronomy, Wayne State University, Detroit, Michigan 48201, USA}
\author{Claude A. Pruneau}
\affiliation{Department of Physics and Astronomy, Wayne State University, Detroit, Michigan 48201, USA}

\date{\today}

\begin{abstract}
  Two-particle number and transverse momentum differential correlation functions are powerful tools for unveiling the detailed dynamics and particle production mechanisms involved in relativistic heavy-ion collisions. 
  Measurements of transverse momentum correlators $P_2$ and $G_2$, in particular, provide added information not readily accessible with better known number correlation functions $R_2$. However, it is found that the $R_2$ and $G_2$ correlators are somewhat  sensitive to  the details of the experimental procedure used to measure them. They  exhibit, in particular,  a dependence on the collision
  centrality bin width, which may have a rather detrimental impact on their physical interpretation.
  A technique to correct these correlators  for collision centrality bin-width averaging is presented.   The technique is based on the hypothesis that  the shape of single- and pair- probability densities vary slower with collision centrality than the corresponding integrated yields. The technique is tested with Pb-Pb simulations based on  the HIJING and ultrarelativistic quantum molecular dynamics models and shown to enable a precision better than 1\% for particles in the kinematic range $0.2 \le p_{\rm T} \le 2.0$ \gevc.
\end{abstract}
\maketitle

\section{Introduction}
\label{sec:Introduction}

Measurements of correlation functions  enable  in-depth exploration of particle 
production mechanisms in relativistic heavy-ion collisions (HIC). Together with 
measurements of the nuclear modification factor, $R_{AA}$, measured number 
correlation functions have provided strong evidence for the formation of a dense 
and opaque medium, a quark gluon plasma (QGP), in the midst of high energy  
heavy-ion 
collisions~\cite{RHICqgp1,RHICqgp2,RHICqgp3,RHICqgp4,LHCqgp1,LHCqgp3,LHCqgp6, 
LHCqgp2,ALICE-PRL2011}. Measurements of transverse momentum differential 
correlation functions have also been added to the experimental 
toolset~\cite{monica,monica_star}. For these as for  differential number 
correlation functions, precision measurements require to account for a number 
of experimental conditions and artifacts~\cite{Ravan:2013lwa}. In this paper, the sensitivity of two-particle differential correlators   to the protocol
used to account for centrality bin width averaging is studied, and  correction
techniques to account for finite collision centrality bins used in typical HIC analyses are considered.

Whether measuring two-particle number or transverse momentum differential correlation functions towards the 
study of collective behavior, the elucidation of particle production mechanisms,  or the analysis of collision systems' evolution towards equilibrium, most data analyses are typically limited by the available statistics. Analyses are thus carried out
with somewhat wide collision centrality bins and aim to report the evolution of the observables with centrality while
accounting for the limited number of events.  It is well established~\cite{RHICqgp1,RHICqgp2,RHICqgp3,RHICqgp4,LHCqgp1,LHCqgp3,LHCqgp6, 
LHCqgp2,ALICE-PRL2011} that both the shape and strength of  two-particle correlations evolve
with collision centrality, i.e., the number, $N_w$, of nucleons wounded  in a given  collision.  Normalized two-particle 
cumulant based correlators such as $R_2$, $P_2$, and $G_2$, defined below,   scale as the inverse of the
number of wounded $N_w$, or the number of nucleon participants, for collisions involving no collectivity and no rescattering of secondaries.
Both the amplitude and the shape of these correlators thus indeed evolve with collision centrality. 
However, measurements of correlation functions are often carried out using fairly wide centrality bins amounting to
5\%, 10\%, or even larger values of the total interaction cross section. Effectively, if events within a given centrality bin 
are analyzed indiscriminately, measured values of the correlator $R_2$, $P_2$, and $G_2$ will amount to
some average across the width of the centrality bins. The issue arises, however, that such averages may not 
be properly calculated unless one explicitly accounts for the fact these three correlators are ratios of quantities that might behave differently across the bins. The mean value theorem, written as
\be\label{eq:meanValueTh}
f(x_0)  = \frac{\int_a^b f(x) dx}{b-a},
\ee   
stipulates that the mean value of a function across an interval $[a,b]$ is equal to the value of the function evaluated at $x_0$, a value of $x$ within the interval $[a,b]$. In general $x_0$ does not correspond to the center of the bin. As will be shown, the correlator $R_2$ is a ratio of a two-particle density by the square (essentially) of a single particle density. Both the single and pair densities are functions of centrality. Evaluation of the average of these densities across a centrality 
bin involves integrals of the form given by Eq.~(\ref{eq:meanValueTh}). The two functions are evaluated across the same interval but may feature significantly different dependence on the centrality, i.e., variable $x$ in Eq.~(\ref{eq:meanValueTh}). The two estimated densities end up corresponding to different values of $x$ (centrality). The $R_2$ correlator, if evaluated as a ratio of these integrals, then amounts to a ratio of quantities 
evaluated at two distinct and unknown values of $x$. It is thus intrinsically biased and does not constitute a proper 
estimator of the average of $R_2$ across the centrality bin. A correction must thus be made to account for or suppress this bias. It will be shown that similar considerations also apply to the correlators $P_2$ and $G_2$.
In addition to the width of the centrality bin, the magnitude of the corrections might  also depend on the rate at which these correlators evolve with collision centrality. It is thus useful to invoke existing heavy-ion collision  models to obtain quasirealistic  correlation functions and simulate their evolution with collision centrality. These
can then be used to assess the magnitude of the corrections and systematic uncertainties associated with 
such corrections.

Correction procedures similar to those presented in this work were used in prior studies but were never formally published. The STAR collaboration used the correction procedure towards 
the study of  differential two-particle transverse momentum  correlations in Au-Au collisions~\cite{Adams:2005ka}, whereas  the ALICE collaboration for measurements of the $R_2$ and
$P_2$ correlators  in $p$-Pb and Pb-Pb collisions~\cite{Abelev:2014ckr}.  
This work establishes a formal record and documentation of these procedures and presents an
estimate of their accuracy based on HIJING and UrQMD, two semirealistic heavy-ion collisions models.

This article is organized as follows. 
Inclusive and conditional particle densities are defined in 
Sec.~\ref{sec:Observable definitions}. The impact of the centrality bin width and a correction formula are first derived for normalized two-particle cumulants (two-particle number differential correlations) in 
Sec.~\ref{sec:Normalized Cumulants}. The study is then extended to 
momentum correlation functions in  Sec.~\ref{sec:TransverseMomentumCorrelators}. 
In Sec.~\ref{sec:Simulations}, estimates of the precision  of the 
method based on simulations carried out with the HIJING~\cite{PhysRevD.44.3501} 
and ultrarelativistic quantum molecular dynamics (UrQMD)~\cite{Bass:1998ca,Bleicher:1999xi} particle event generators are presented. This 
work is summarized in Sec.~\ref{sec:Summary}.

\section{Observable definitions}
\label{sec:Observable definitions}

Whether analyzing  $pp$, $p$-Pb, or $A$-$A$ collisions, it is legitimate to classify events on the basis of the energy, $E_{\rm ref}$, or the charged particle multiplicity, $m$, measured in a  reference acceptance $\Omega_{\rm ref}$. 
 Studies~\cite{Rogly:2018ddx} have shown that in  $A$-$A$ collisions, this multiplicity maps rather narrowly onto 
 the impact parameter $b$ of the collisions, particularly in the case of mid- to central-collisions. While such narrow mapping does not exist for $p$-A or $pp$ interactions, it remains appropriate to classify 
 collisions based on the multiplicity  $m$ as it provides some measure of the momentum transfer and
 projectile kinetic energy dissipated in the collisions. In general, one indeed finds that the single and pair density
 measured in the fiducial acceptance $\Omega_{\rm ref}$ are functions of the reference multiplicity $m$. Thus single-  and two-particle conditional densities (i.e., for a given $m$) are defined as
\begin{widetext}
\be
\rho_1(\eta,\varphi,p_{\rm T}|m)& =& \left. \frac{1}{p_{\rm T} }\frac{dN}{d\varphi d\eta dp_{\rm T}  } \right|_m\text{,}\\
\rho_2(\eta_1,\varphi_1,p_{\rm T,1},\eta_2,\varphi_2,p_{\rm T,2}|m) &=& \left. \frac{1}{p_{\rm T,1}p_{\rm T,2}}\frac{d^2N}{d\varphi_1 d\eta_1 dp_{\rm T,1}d\varphi_2 d\eta_2 dp_{\rm T,2}   } \right|_m\text{.}
\ee
\end{widetext}
Typically, correlation analyses are carried out over fixed $p_{\rm T}$ ranges $[p_{\rm T,\min},p_{\rm T,\max}]$. One thus also defines conditional single- and two-particle densities integrated over 
$p_{\rm T}$ as 
\begin{widetext}
\be
\label{eq:rho1m}
\rho_1(\eta,\varphi|m)& =& \int  \rho_1(\eta,\varphi,p_{\rm T}|m) dp_{\rm T}\text{,} \\
\label{eq:rho2m}
\rho_2(\eta_1,\varphi_1,\eta_2,\varphi_2|m) &=&\int \rho_2(\eta_1,\varphi_1,p_{\rm T,1},\eta_2,\varphi_2,p_{\rm T,2}|m) dp_{\rm T,1} dp_{\rm T,2} \text{,}
\ee
\end{widetext}
where (here and in the following) it is understood that the integrals in $p_{\rm T}$ are taken over the fixed range $[p_{\rm T,\min},p_{\rm T,\max}]$.
In practice, it is usually not possible (or meaningful) to measure the densities $\rho_1(\eta,\varphi|m)$ and 
$\rho_2(\eta_1,\varphi_1,\eta_2,\varphi_2|m)$ for unit resolution in $m$. One must then evaluate the densities
within finite width bins of multiplicity $[m_{\min,k},m_{\max,k}]$ (where $k=1,\ldots, K$, represents one of  $K$ ``centrality" bins used in the analysis) as weighted average of the densities 
across the bins according to
\begin{widetext}
\be
\label{eq:rho1mK}
\bar \rho_1^{(k)}(\eta,\varphi)& =& \frac{1}{Q_k} \sum_{m=m_{\min,k}}^{m_{\max,k}} q(m)  \rho_1(\eta,\varphi|m), \\
\label{eq:rho2mK}
\bar \rho_2^{(k)}(\eta_1,\varphi_1,\eta_2,\varphi_2) &=& \frac{1}{Q_k} \sum_{m=m_{\min,k}}^{m_{\max,k}} q(m)  \rho_2(\eta_1,\varphi_1,\eta_2,\varphi_2|m)
\ee
\end{widetext}
with 
\be 
Q_k =  \sum_{m=m_{\min,k}}^{m_{\max,k}} q(m)
\ee
and $q(m)$ representing the probability of events with multiplicity $m$ in the reference acceptance.

\section{Normalized Cumulants: $R_2$}
\label{sec:Normalized Cumulants}

Two-particle normalized cumulants~\cite{Ravan:2013lwa}  are defined according to 
\begin{widetext}
\be
\label{eq:R2}
R_2(\eta_1,\varphi_1,\eta_2,\varphi_2)& =& \frac{\rho_2(\eta_1,\varphi_1,\eta_2,\varphi_2) - \rho_1(\eta_1,\varphi_1)\rho_1(\eta_2,\varphi_2)}{\rho_1(\eta_1,\varphi_1)\rho_1(\eta_2,\varphi_2)}
\ee
\end{widetext}
and  form the basis of many two-particle correlation analyses. For a recent study on the evolution with centrality of the two-particle differential number density correlation $R_2$ in $p$-Pb and Pb-Pb collisions, by the ALICE collaboration see Ref.~\cite{Acharya:2018ddg}. Experimentally, in the study of $A$-$A$ collisions, 
  it is common to average $R_2$ across large collision centrality bins. Given the width of such bins may depend on the specificities 
  of a given experiment (e.g., the size of the reference acceptance), normalized cumulants $R_2$ are thus bin-width dependent. A correction procedure is then required to account for the finite width 
of
  the centrality bins. 
  
  In principle, one would like to set  the collision centrality bins in terms of collision impact parameter ranges (alternatively the number of 
  wounded nucleons). In practice, only proxies of the impact parameter are at best possible and one usually expresses  the bins directly in terms of such proxies or in terms of the fractional 
cross section measured within a centrality (multiplicity) bin. Let $m$, the multiplicity measured
  in the reference acceptance $\Omega_{\rm ref}$ represent such a proxy. As for densities, one can thus define conditional normalized cumulants according to 
\begin{widetext}
\be
\label{eq:R2m}
R_2(\eta_1,\varphi_1,\eta_2,\varphi_2|m)& =& \frac{\rho_2(\eta_1,\varphi_1,\eta_2,\varphi_2|m)}{\rho_1(\eta_1,\varphi_1|m)\rho_1(\eta_2,\varphi_2|m)} - 1.
\ee 
\end{widetext}
However, measurements of $R_2(\eta_1,\varphi_1,\eta_2,\varphi_2|m)$ for $m$ unit resolution  are typically not possible due to limitations associated
with the finite size of datasets, CPU time, or storage considerations. It is also impractical to report differential correlators $R_2(\eta_1,\varphi_1,\eta_2,\varphi_2|m)$ for a very large number of 
values of $m$. In practice, one thus seeks to report correlators $R_2$ averaged over reference multiplicity bins $[m_{\min,k},m_{\max,k}]$. As for averages of densities defined in 
Sec.~\ref{sec:Observable definitions}, this is nominally achieved according to 
\begin{widetext}
\be
\label{eq:R2mK}
\bar R_2^{(k)}(\eta_1,\varphi_1,\eta_2,\varphi_2) &=& \frac{1}{Q_k} \sum_{m=m_{\min,k}}^{m_{\max,k}} q(m) R_2(\eta_1,\varphi_1,\eta_2,\varphi_2|m),
\ee
\end{widetext}
where $q(m)$ represents the probability of events with multiplicity $m$ in the reference acceptance.
One may then determine $R_2^{(k)}$ according to 
\begin{widetext}
\be
\label{eq:R2mKBin}
R_2^{({\rm Bin},k)}(\eta_1,\varphi_1,\eta_2,\varphi_2) &=& \frac{ \rho_2^{({\rm Bin},k)}(\eta_1,\varphi_1,\eta_2,\varphi_2) }{\rho_1^{({\rm Bin},k)}(\eta_1,\varphi_1)\rho_1^{({\rm 
Bin},k)}(\eta_2,\varphi_2)} - 1\text{,}
\ee
\end{widetext}
in which $\rho_1^{({\rm Bin},k)}$ and $\rho_2^{({\rm Bin},k)}$ are single and pair densities measured directly using finite width bins in $m$. In the absence of biases or event detection 
inefficiencies, one has
\be
\rho_1^{({\rm Bin},k)}(\eta_1,\varphi_1) &=& \bar \rho_1^{(k)}(\eta,\varphi), \\ 
\rho_2^{({\rm Bin},k)}(\eta_1,\varphi_1,\eta_2,\varphi_2) &=& \bar \rho_2^{(k)}(\eta_1,\varphi_1,\eta_2,\varphi_2).
\ee
The quantities $R_2^{({\rm Bin},k)}$ and  $\bar R_2^{(k)}$ are thus clearly distinct:
\begin{widetext}
\be
\label{eq:R2mKandBin}
R_2^{({\rm Bin},k)}(\eta_1,\varphi_1,\eta_2,\varphi_2) &=& 
\frac{ \frac{1}{Q_k} \sum_{m=m_{\min,k}}^{m_{\max,k}} q(m)  \rho_2(\eta_1,\varphi_1,\eta_2,\varphi_2|m)}
{
\left[\frac{1}{Q_k} \sum_{m=m_{\min,k}}^{m_{\max,k}} q(m)  \rho_1(\eta_1,\varphi_1|m)\right]
\left[\frac{1}{Q_k} \sum_{m'=m_{\min,k}}^{m_{\max,k}} q(m')  \rho_1(\eta_2,\varphi_2|m')\right]} - 1\text{,} \\ 
\bar R_2^{(k)}(\eta_1,\varphi_1,\eta_2,\varphi_2) &= & \left[\frac{1}{Q_k}\sum_{m=m_{\min,k}}^{m_{\max,k}} q(m)  \frac{\rho_2(\eta_1,\varphi_1,\eta_2,\varphi_2|m)}
{\rho_1(\eta_1,\varphi_1|m) \rho_1(\eta_2,\varphi_2|m)} \right] - 1 \text{.}
\ee
\end{widetext}
To quantify the difference, the particle densities are written in terms of single- and two-particle probability densities,  $\PDF_1(\eta,\varphi|m)$ and $\PDF_2(\eta_1,\varphi_1,\eta_2,\varphi_2|m)$, according to 
\begin{widetext}
\be\label{eq:pdf1}
\rho_1(\eta,\varphi|m)& =& \la n\ra_m \PDF_1(\eta,\varphi|m), \\
\label{eq:pdf2}
\rho_2(\eta_1,\varphi_1,\eta_2,\varphi_2|m) &=& \la n(n-1)\ra_m \PDF_2(\eta_1,\varphi_1,\eta_2,\varphi_2|m),
\ee
\end{widetext}
where, by definition, $\PDF_1$ and $\PDF_2$ respectively satisfy 
\be\label{eq:pdf1Int}
\int d\varphi d\eta \, \PDF_1(\eta,\varphi|m) = 1\text{,}\\
\label{eq:pdf2Int}
\int d\varphi_1 d\eta_1 d\varphi_2 d\eta_2 \,  \PDF_2(\eta_1,\varphi_1,\eta_2,\varphi_2|m) = 1\text{,}
\ee
as probability densities. 
The quantities 
$\la n\ra_m$ and $\la n(n-1)\ra_m$ are the  mean number of particles and the mean number of pairs of particles in the acceptance of the measurement at a given reference multiplicity $m$;  
$\PDF_1(\eta,\varphi|m)$ is the probability of finding a particle at $\eta,\varphi$ when the multiplicity is $m$, and 
$\PDF_2(\eta_1,\varphi_1,\eta_2,\varphi_2|m)$ is the joint probability of measuring particles at $\eta_1,\varphi_1$ and $\eta_2,\varphi_2$ when the multiplicity is $m$.  In general, one expects $\la 
n\ra_m$ and $\la n(n-1)\ra_m$ to scale approximately linearly and quadratically, respectively, with $m$. Let us assume that the shape of $\PDF_1(\eta,\varphi|m)$ and 
$\PDF_2(\eta_1,\varphi_1,\eta_2,\varphi_2|m)$ 
change little through a centrality bin. One can then write 
\begin{widetext}
\be
R_2^{({\rm Bin},k)}(\eta_1,\varphi_1,\eta_2,\varphi_2)&=&\alpha \,  \frac{\bar \PDF_2(\eta_1,\varphi_1,\eta_2,\varphi_2)}{\bar \PDF_1(\eta_1,\varphi_1)\bar \PDF_1(\eta_2,\varphi_2)} -1
\ee
\end{widetext}
with 
\be\label{eq:alpha}
\alpha = \frac{\frac{1}{Q_k} \sum_{m=m_{\min,k}}^{m_{\max,k}} q(m) \la n(n-1)\ra_m}
{\left(\frac{1}{Q_k} \sum_{m=m_{\min,k}}^{m_{\max,k}} q(m) \la n\ra_m \right)^2}
\ee
and $\bar \PDF_1(\eta_1,\varphi_1)$ and $\bar \PDF_2(\eta_1,\varphi_1,\eta_2,\varphi_2)$ are the bin-width 
averaged values of the single- and two-particle probability densities.
Similarly, one can also write 
\begin{widetext}
\be
\bar R_2^{(k)}(\eta_1,\varphi_1,\eta_2,\varphi_2) &=& \beta \, \frac{  \bar \PDF_2(\eta_1,\varphi_1,\eta_2,\varphi_2) }{  \bar \PDF_1(\eta_1,\varphi_1)  \bar \PDF_1(\eta_2,\varphi_2)} - 1
\ee
\end{widetext}
with 
\be \label{eq:beta}
\beta  = \frac{1}{Q_k} \sum_{m=m_{\min,k}}^{m_{\max,k}} q(m)
 \frac{ \la n(n-1)\ra_m  }{\la n\ra_m^2  }.
\ee
One thus finds that for sufficiently narrow bins (such that $\PDF_1$ and $\PDF_2$ can be considered approximately 
invariant within)
\begin{widetext}
\be\label{eq:r2corr}
\bar R_2^{(k)}(\eta_1,\varphi_1,\eta_2,\varphi_2)  = \beta \alpha^{-1} 
\left( R_2^{({\rm Bin},k)}(\eta_1,\varphi_1,\eta_2,\varphi_2)+1 \right) - 1
\ee
\end{widetext}

The above expression assumes that the event detection efficiency is either unity or constant across a bin $k$. If the 
event detection efficiency, $\varepsilon(m)$, varies across  $[m_{\min,k},m_{\max,k}]$, and the densities $\PDF_1$ 
and $\PDF_2$ are not biased by this dependence, then the coefficients $\alpha$ and $\beta$
may be written as 
\be\label{eq:alpha2}
\alpha &=& \frac{\frac{1}{Q_k} \sum_{m=m_{\min,k}}^{m_{\max,k}} \frac{q^{*}(m)}{ \varepsilon(m)}  \la n(n-1)\ra_m}
{\left(\frac{1}{Q_k} \sum_{m=m_{\min,k}}^{m_{\max,k}} \frac{q^{*}(m)}{ \varepsilon(m)} \la n\ra_m \right)^2}, \\
\label{eq:beta2}
\beta  &=& \frac{1}{Q_k} \sum_{m=m_{\min,k}}^{m_{\max,k}} \frac{q^{*}(m)}{ \varepsilon(m)}
  \frac{ \la n(n-1)\ra_m  }{\la n\ra_m^2  },
\ee
where $q^{*}(m)$ is now the observed (uncorrected) probability distribution of multiplicity $m$.

\section{Transverse Momentum Correlators}
\label{sec:TransverseMomentumCorrelators}

Several forms of transverse momentum correlators have been proposed and 
reported in the recent 
literature~\cite{Adams:2005aw,S.GavinAPHA:2006Diffusion,monica,monica_star,
Gavin:2016jfw,Gavin:2016hmv,Adam:2017ucq, Acharya:2018ddg}. 
In this work, the focus is on the $G_2$ correlator proposed by Gavin \textit{et 
al}.~\cite{S.GavinAPHA:2006Diffusion,Gavin:2016jfw,Gavin:2016hmv,monica_star} to 
study transverse momentum current correlations and the $P_2$ correlator designed 
to be sensitive to transverse momentum 
fluctuations~\cite{monica,Adam:2017ucq,Acharya:2018ddg}. 

\subsection{$G_2$ correlator}
\label{sec:G2Correlators}

At fixed reference multiplicity, $m$, the differential transverse momentum 
correlator $G_2$ can be written as~\cite{S.GavinAPHA:2006Diffusion}
\begin{widetext}
\be \label{eq:G2Def}
  G_2 (\eta_1,\varphi_1,\eta_2,\varphi_2|m) 
      &=& \frac{ \int dp_{\rm T,1}  \int dp_{\rm T,2} \, p_{\rm T,1} \, p_{\rm T,2} 
                  \, \rho_2(\eta_1,\varphi_1,p_{\rm T,1},\eta_2,\varphi_2,p_{\rm T,2}|m) }
               { \rho_1(\eta_1,\varphi_1|m) \, \rho_1(\eta_2,\varphi_2|m) } \\ \nonumber
 & & - \la p_{\rm T}(\eta_1,\varphi_1|m)\ra\la p_{\rm T}(\eta_2,\varphi_2|m)\ra\text{,}
\ee
\end{widetext}
where
\be 
\la p_{\rm T}(\eta,\varphi|m)\ra &=&  \frac{ \int dp_{\rm T}   p_{\rm T} \, \rho_1(\eta,\varphi,p_{\rm T}|m) }
                 { \rho_1(\eta,\varphi|m) }
\ee
is the event inclusive average particle transverse momentum at $\eta$, $\varphi$.
Evaluation of the centrality-bin averaged  $\bar G_2^{(k)}(\eta_1,\varphi_1,\eta_2,\varphi_2)$ proceeds as 
for $R_2$ and one writes
\begin{widetext}
\be
\bar G_2^{(k)}(\eta_1,\varphi_1,\eta_2,\varphi_2) =  \frac{1}{Q_k} \sum_{m=m_{\min,k}}^{m_{\max,k}} q(m) \,G_2(\eta_1,\varphi_1,\eta_2,\varphi_2|m).
\ee
\end{widetext}
Introducing 
\begin{widetext}
\be
\label{eq:s1m}
S_1(\eta,\varphi|m)  &=& \int p_{\rm T} \, dp_{\rm T} \,  \rho_1(\eta,\varphi,p_{\rm T}|m),\\ 
&=& \la n\ra_m \, \PDF_1^{p_{\rm T}}(\eta,\varphi|m),\\
\label{eq:s2m}
S_2(\eta_1,\varphi_1,\eta_2,\varphi_2|m)  &=& \int p_{\rm T,1} \, dp_{\rm T,1} \int  p_{\rm T,2}  dp_{\rm T,2} \,  
                   \rho_2(\eta_1,\varphi_1,p_{\rm T,1},\eta_2,\varphi_2,p_{\rm T,2}|m), \\ 
&=&  \la n(n-1) \ra_m \, \PDF_2^{p_{\rm T}p_{\rm T}}(\eta_1,\varphi_1,\eta_2,\varphi_2|m)
\ee
\end{widetext}
with 
\begin{widetext}
\be
\label{eq:P1pT}
\PDF_1^{p_{\rm T}}(\eta,\varphi|m) &=& \int p_{\rm T} \, dp_{\rm T} \,  \PDF_1(\eta,\varphi,p_{\rm T}|m), \\ 
\label{eq:P2pTpT}
\PDF_2^{p_{\rm T}p_{\rm T}}(\eta_1,\varphi_1,\eta_2,\varphi_2|m) &=&
 \int p_{\rm T,1} \,dp_{\rm T,1}  \int p_{\rm T,2} \,  dp_{\rm T,2} \,  
          \PDF_2(\eta_1,\varphi_1,p_{\rm T,1},\eta_2,\varphi_2,p_{\rm T,2}|m),
\ee
\end{widetext}
Eq.~(\ref{eq:G2Def}) is written as
\begin{widetext}
\be 
  G_2 (\eta_1,\varphi_1,\eta_2,\varphi_2|m) &=& \frac{S_2(\eta_1,\varphi_1,\eta_2,\varphi_2|m)}{ \rho_1(\eta_1,\varphi_1|m) \, \rho_1(\eta_2,\varphi_2|m) }  - 
\frac{S_1(\eta_1,\varphi_1|m)}{\rho_1(\eta_1,\varphi_1|m)}\frac{S_1(\eta_2,\varphi_2|m)}{\rho_1(\eta_2,\varphi_2|m)} \\ 
      &=& \frac{\la n(n-1) \ra_m}{\la n \ra_m^2} \frac{ \PDF_2^{p_{\rm T}p_{\rm T}}(\eta_1,\varphi_1,\eta_2,\varphi_2|m) }
               { \PDF_1(\eta_1,\varphi_1|m) \, \PDF_1(\eta_2,\varphi_2|m) } 
 -  \frac{\PDF_1^{p_{\rm T}}(\eta_1,\varphi_1|m)}{\PDF_1(\eta_1,\varphi_1|m)}  \frac{\PDF_1^{p_{\rm T}}(\eta_2,\varphi_2|m)}{\PDF_1(\eta_2,\varphi_2|m)}\text{.}
\ee
\end{widetext}
If the ratios $\PDF_1^{p_{\rm T}}/\PDF_1$ and $\PDF_2^{p_{\rm T}p_{\rm T}}/\PDF_1\PDF_1$ have a modest dependence on the reference multiplicity (within a bin $k$),
then it is legitimate to replace them by averages and one gets
\begin{widetext}
\be
  G_2 (\eta_1,\varphi_1,\eta_2,\varphi_2|m) &=& \frac{\la n(n-1) \ra_m}{\la n \ra_m^2} \frac{ \bar \PDF_2^{p_{\rm T}p_{\rm T} (k)}(\eta_1,\varphi_1,\eta_2,\varphi_2) }
               { \bar \PDF_1^{ (k)}(\eta_1,\varphi_1) \, \bar \PDF_1^{ (k)}(\eta_2,\varphi_2) } 
  -  \frac{\bar \PDF_1^{p_{\rm T} (k)}(\eta_1,\varphi_1)}{\bar \PDF_1^{(k)}(\eta_1,\varphi_1)}  \frac{\bar \PDF_1^{p_{\rm T} (k)}(\eta_2,\varphi_2)}{\bar \PDF_1^{(k)}(\eta_2,\varphi_2)}\text{.}
\ee
\end{widetext}
The centrality-bin averaged correlator $\bar 
G_2^{(k)}(\eta_1,\varphi_1,\eta_2,\varphi_2)$ may then be written as
\begin{widetext}
\be
\bar G_2^{(k)}(\eta_1,\varphi_1,\eta_2,\varphi_2) = \beta \frac{ \bar \PDF_2^{p_{\rm T}p_{\rm T} (k)}(\eta_1,\varphi_1,\eta_2,\varphi_2) }
               { \bar \PDF_1^{ (k)}(\eta_1,\varphi_1) \, \bar \PDF_1^{ (k)}(\eta_2,\varphi_2) } - 
               \frac{\bar \PDF_1^{p_{\rm T}(k)}(\eta_1,\varphi_1)}{\bar \PDF_1^{ (k)}(\eta_1,\varphi_1)}  
               \frac{\bar \PDF_1^{p_{\rm T}(k)}(\eta_2,\varphi_2)}{\bar \PDF_1^{ (k)}(\eta_2,\varphi_2)}
\ee
\end{widetext}
with $\beta$ defined as in Eq.~(\ref{eq:beta}). However, if it is not possible to carry out the analysis in fine (unit) bins  of $m$, the numerators and denominators of $G_2$ must be separately averaged over the range $[m_{\min,k},m_{\max,k}]$. 
Assuming  the ratios $\PDF_1^{p_{\rm T}}/\PDF_1$ and $\PDF_2^{p_{\rm T}p_{\rm T}}/\PDF_1\PDF_1$ have a modest dependence on the reference multiplicity, one   gets 
\begin{widetext}
\be\label{eq:G2Factor}
G_2^{({\rm Bin},k)}(\eta_1,\varphi_1,\eta_2,\varphi_2) = \alpha \frac{ \bar \PDF_2^{p_{\rm T}p_{\rm T} (k)}(\eta_1,\varphi_1,\eta_2,\varphi_2) }
               { \bar \PDF_1^{ (k)}(\eta_1,\varphi_1) \, \bar \PDF_1^{ (k)}(\eta_2,\varphi_2) } - 
               \frac{\bar \PDF_1^{p_{\rm T}(k)}(\eta_1,\varphi_1)}{\bar \PDF_1^{ (k)}(\eta_1,\varphi_1)}  
               \frac{\bar \PDF_1^{p_{\rm T}(k)}(\eta_2,\varphi_2)}{\bar \PDF_1^{ (k)}(\eta_2,\varphi_2)}
\ee
\end{widetext}
with $\alpha$ defined as in Eq.~(\ref{eq:alpha}). Identifying $\la p_{\rm T}(\eta,\varphi)\ra^{(\text{Bin},k)} = \bar \PDF_1^{p_{\rm T}(k)}(\eta,\varphi)/\bar \PDF_1^{(k)}(\eta,\varphi)$, one finally gets that the 
desired correlator $\bar G_2^{(k)}(\eta_1,\varphi_1,\eta_2,\varphi_2)$ may be determined as 
\begin{widetext}
\begin{align}
\bar{G}_2^{(k)}(\eta_1,\varphi_1,\eta_2,\varphi_2) = &\, {\beta}{\alpha^{-1}} \left(G_2^{({\rm Bin},k)}(\eta_1,\varphi_1,\eta_2,\varphi_2) + 
                                                      \la p_{\rm T}(\eta_1,\varphi_1)\ra^{({\rm Bin},k)} \la p_{\rm T}(\eta_2,\varphi_2)\ra^{({\rm Bin},k)} \right) \nonumber \\
                                                  &\quad - \la p_{\rm T}(\eta_1,\varphi_1)\ra^{({\rm Bin},k)} \la p_{\rm T}(\eta_2,\varphi_2)\ra^{({\rm Bin},k)} \label{eq:g2corr}, \\
                                                  = & {\beta}{\alpha^{-1}}  G_2^{({\rm Bin},k)}(\eta_1,\varphi_1,\eta_2,\varphi_2)  + \left({\beta}{\alpha^{-1}} -1 \right) \la p_{\rm T}(\eta_1,\varphi_1)\ra^{({\rm Bin},k)} \la p_{\rm T}(\eta_2,\varphi_2)\ra^{({\rm Bin},k)}\text{.}\nonumber                                                  
\end{align}
\end{widetext}

\subsection{$P_2$ correlator}
\label{sec:P2Correlators}

The $P_2$ correlator  is defined as 
\begin{widetext}
\be
\label{eq:P2m}
P_2(\eta_1,\varphi_1,\eta_2,\varphi_2|m) &=& \frac{1}{\la p_{\rm T}\ra^2} 
\frac{\int \Delta p_{\rm T,1} dp_{\rm T,1} \, \int \Delta p_{\rm T,2} dp_{\rm 
T,2} \,  \rho_2(\eta_1,\varphi_1,p_{\rm T,1},\eta_2,\varphi_2,p_{\rm 
T,2}|m)}{\rho_2(\eta_1,\varphi_1,\eta_2,\varphi_2|m)}\text{,}
\ee
\end{widetext}
where $\Delta p_{\rm T,i} \equiv p_{\rm T,i} - \la p_{\rm T}\ra$ and the $\la 
p_{\rm T}\ra^2$ normalization insures $P_2$ is dimensionless. 
Introducing~(\ref{eq:pdendptdpt}) one  next verifies that $P_2$ is insensitive to multiplicity fluctuations~\cite{monica,Adam:2017ucq,Acharya:2018ddg}:
\begin{widetext}
\be
\label{eq:pdendptdpt}
\PDF_2^{\Delta p_{\rm T}\Delta p_{\rm T}}(\eta_1,\varphi_1,\eta_2,\varphi_2|m) 
&=&
\int \Delta p_{\rm T,1} dp_{\rm T,1}  \int \Delta p_{\rm T,2} dp_{\rm T,2} \,  
\PDF_2(\eta_1,\varphi_1,p_{\rm T,1},\eta_2,\varphi_2,p_{\rm T,2}|m)\text{.}
\ee
\end{widetext}
Factorizing the 
two-particle density as in Eq.~(\ref{eq:pdf2}), one gets
\begin{widetext}
\be
P_2(\eta_1,\varphi_1,\eta_2,\varphi_2|m)& =& \left(\frac{\PDF_1}{\PDF_1^{p_{\rm 
T}}}\right)^2
\frac{\la n(n-1)\ra_m}{\la n(n-1)\ra_m}\frac{\PDF_2^{\Delta p_{\rm T}\Delta 
p_{\rm 
T}}(\eta_1,\varphi_1,\eta_2,\varphi_2|m)}{\PDF_2(\eta_1,\varphi_1,\eta_2,
\varphi_2|m)}\text{.}
\ee
\end{widetext}

If the ratios $\PDF_1^{p_{\rm T}}/\PDF_1$ and $\PDF_2^{\Delta p_{\rm T} \Delta 
p_{\rm T}}/\PDF_2$ have a modest dependence on the reference multiplicity 
(within a bin $k$),
then it is legitimate to replace them by averages and one gets
\begin{widetext}
\be
P_2(\eta_1,\varphi_1,\eta_2,\varphi_2|m)& =& 
\left(\frac{\bar{\PDF}_1^{(k)}}{\bar{\PDF}_1^{p_{\rm T}(k)}}\right)^2
\frac{\bar \PDF_2^{\Delta p_{\rm T}\Delta p_{\rm T} 
(k)}(\eta_1,\varphi_1,\eta_2,\varphi_2)}{\bar 
\PDF_2^{(k)}(\eta_1,\varphi_1,\eta_2,\varphi_2)}\text{,}
\ee
\end{widetext}
which is indeed independent of the number of pairs and thus multiplicity 
fluctuations, since the functions $\bar 
\PDF_2(\eta_1,\varphi_1,\eta_2,\varphi_2|m)$ are probability densities. 
The centrality-bin averaged value of $P_2$ is calculated according to
\begin{widetext}
\be
\bar{P}_2^{(k)}(\eta_1,\varphi_1,\eta_2,\varphi_2)& =& 
\left(\frac{\bar{\PDF}_1^{(k)}}{\bar{\PDF}_1^{p_{\rm T}(k)}}\right)^2
\frac{\bar \PDF_2^{\Delta p_{\rm T}\Delta p_{\rm T} 
(k)}(\eta_1,\varphi_1,\eta_2,\varphi_2)}{\bar 
\PDF_2^{(k)}(\eta_1,\varphi_1,\eta_2,\varphi_2)}\text{,}
\ee
\end{widetext}
while the ratio of bin averages is 
\begin{widetext}
\be
P_2^{({\rm Bin},k)}(\eta_1,\varphi_1,\eta_2,\varphi_2) & =& 
\left(\frac{\bar{\PDF}_1^{(k)}}{\bar{\PDF}_1^{p_{\rm T}(k)}}\right)^2
\frac{\bar \PDF_2^{\Delta p_{\rm T}\Delta p_{\rm T} 
(k)}(\eta_1,\varphi_1,\eta_2,\varphi_2)}{\bar 
\PDF_2^{(k)}(\eta_1,\varphi_1,\eta_2,\varphi_2)}\text{.}
\ee
\end{widetext}
One thus finds that 
\be\label{eq:p2corr}
\bar{P}_2^{(k)}(\eta_1,\varphi_1,\eta_2,\varphi_2) &=& P_2^{({\rm 
Bin},k)}(\eta_1,\varphi_1,\eta_2,\varphi_2)\text{,}
\ee
featuring a unitary correction factor. The  $P_2$ correlator is thus indeed not 
affected by the width of the multiplicity bins so long as the probability 
densities $\PDF_1$ and  $\PDF_2$ exhibit only modest
shape variations within those bins.

\section{Simulations}
\label{sec:Simulations}

\begin{figure}
 \includegraphics[scale=0.50,keepaspectratio=true,clip=true,trim=10pt 20pt 0pt 
10pt]
  {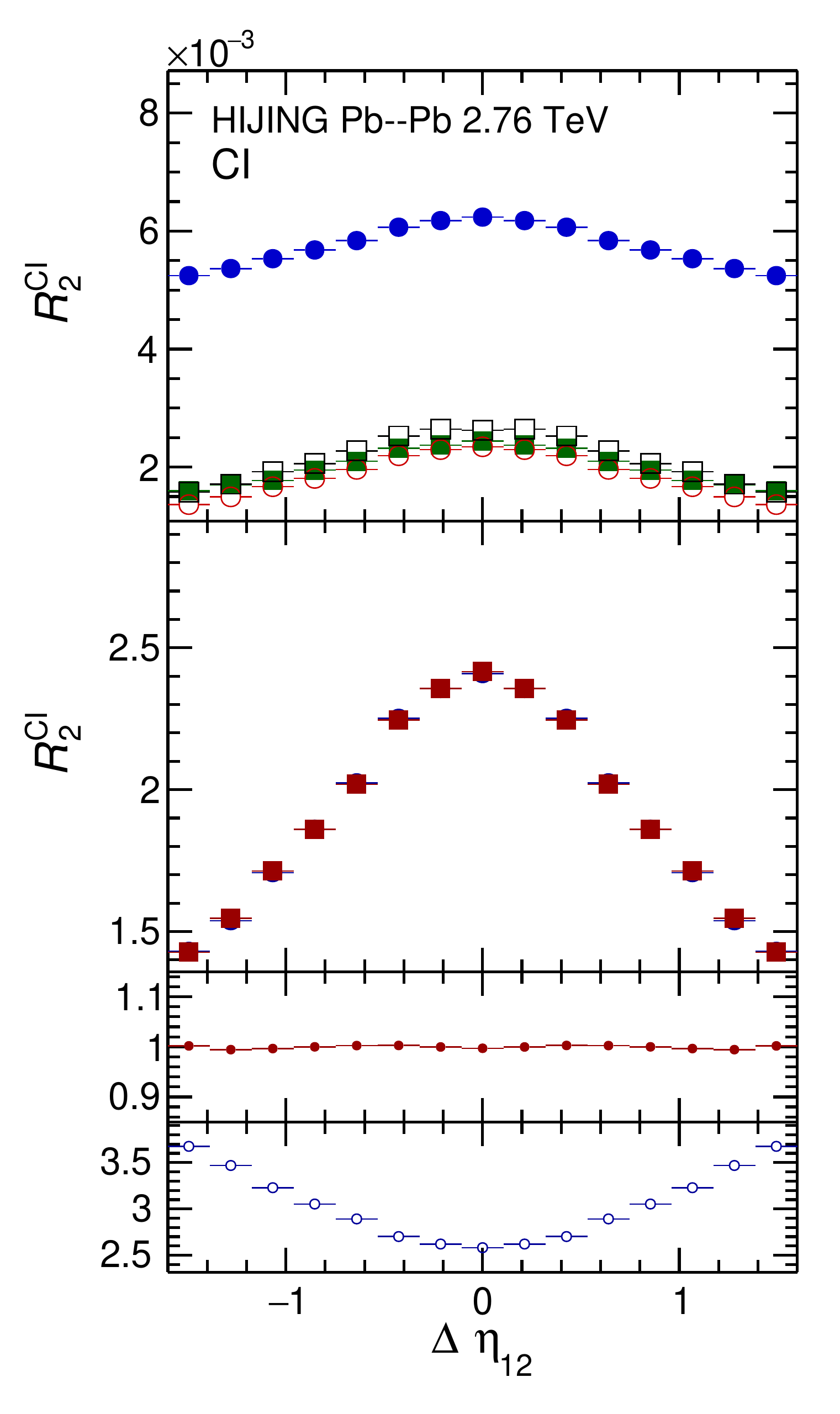}
 \includegraphics[scale=0.50,keepaspectratio=true,clip=true,trim=10pt 20pt 0pt 
10pt]
  {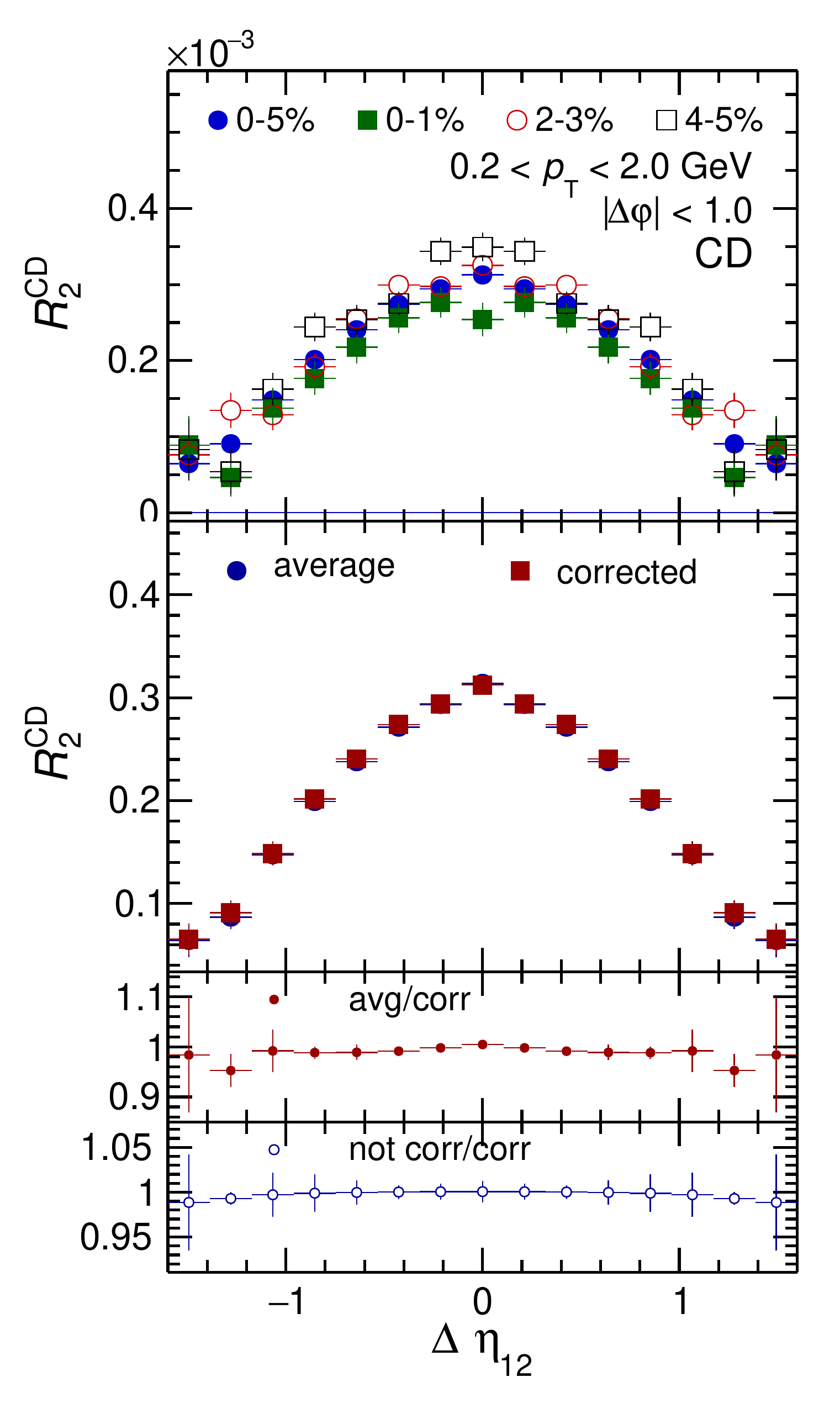} \par
  \hspace{-0.6in}
\caption{\label{fig:r2etahijing}Longitudinal projection of the two-particle
normalized cumulants $\RtwoCI$ (left) and $\RtwoCD$ (right) for the most 
central Pb-Pb collisions produced with the HIJING event generator. Top panels 
show uncorrected results for the 0--5 \% centrality bin together with those from 
the  0--1 \%, 2--3 \%, and 4--5 \% centrality bins, middle panels show 
corrected 0--5 \% centrality bin results compared with the weighted average of 
those from the 0--1 \%, 1--2 \%, 2--3 \%, 3--4 \%, and 4--5 \% centrality bins, while 
bottom panels show their ratio.}
\end{figure}

\begin{figure}
 \includegraphics[scale=0.50,keepaspectratio=true,clip=true,trim=10pt 20pt 0pt 
10pt]
  {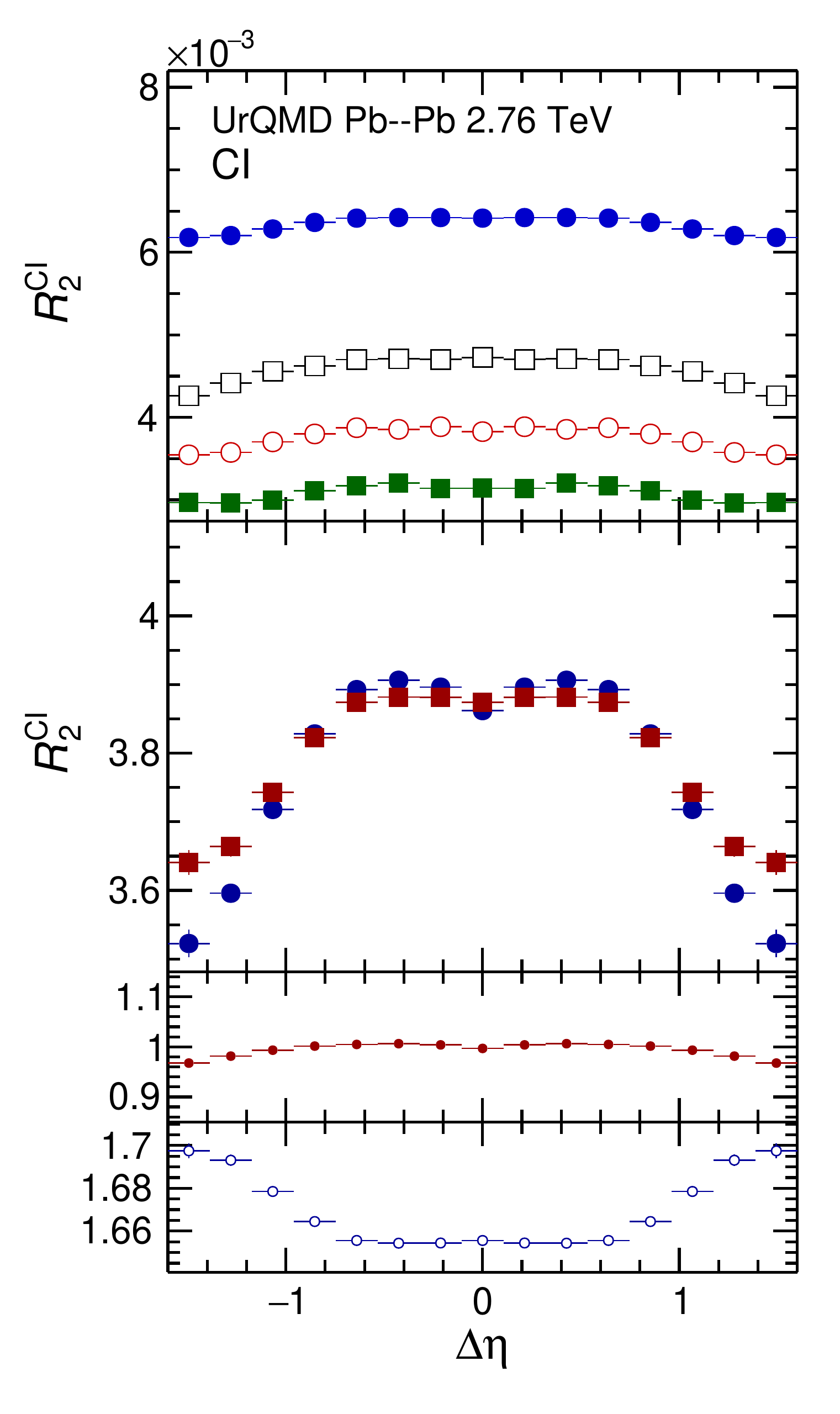}
 \includegraphics[scale=0.50,keepaspectratio=true,clip=true,trim=10pt 20pt 0pt 
10pt]
  {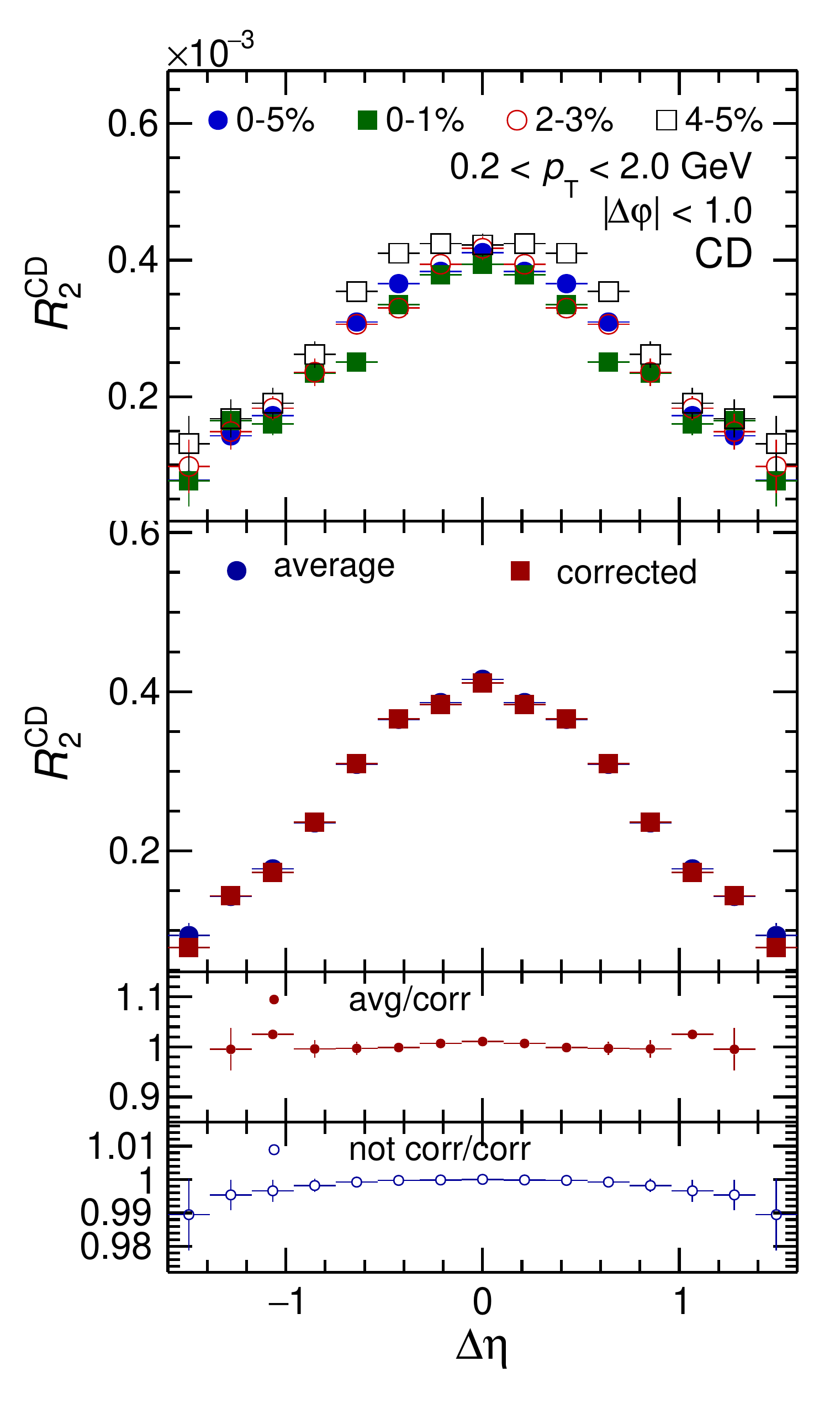} \par
\caption{\label{fig:r2etaurqmd}Longitudinal projection of the two-particle
normalized cumulants $\RtwoCI$ (left) and $\RtwoCD$ (right) for 5\% most 
central Pb-Pb collisions produced with the UrQMD event generator. Top panels 
show uncorrected results for the 0--5 \% centrality bin together with those from 
the 0--1 \%, 2--3 \%, and 4--5 \% centrality bins, middle panels show corrected 
0--5 \% centrality bin results compared with the weighted average of those from 
the 0--1 \%, 1--2 \%, 2--3 \%, 3--4 \%, and 4--5 \% centrality bins, while bottom panels 
show their ratio.}
\end{figure}

\begin{figure*}
 \hspace{-0.4in}
 \includegraphics[scale=0.40,keepaspectratio=true,clip=true,trim=10pt 20pt 22pt 
10pt]
  {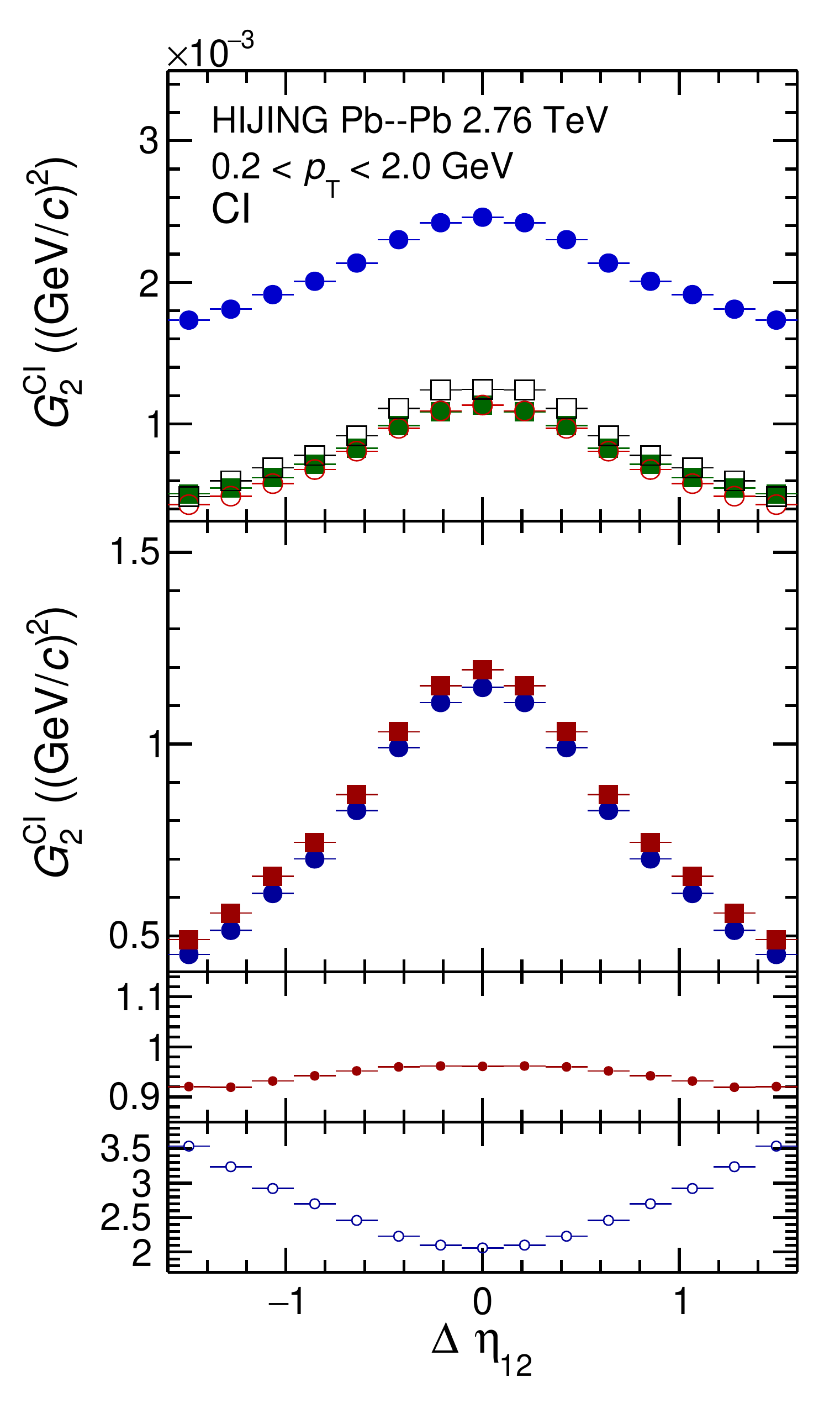}
 \includegraphics[scale=0.40,keepaspectratio=true,clip=true,trim=10pt 20pt 22pt 
10pt]
  {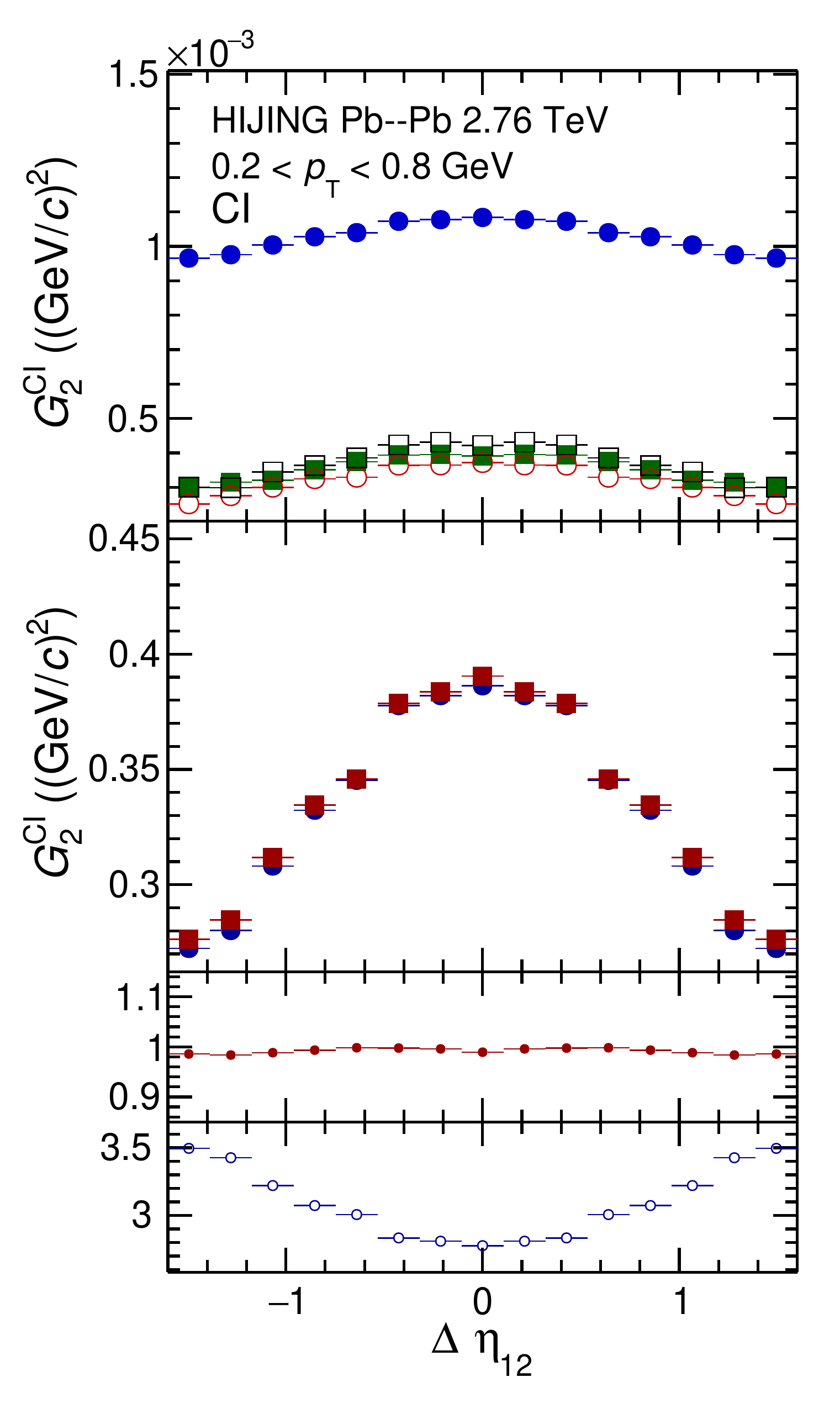}
 \includegraphics[scale=0.40,keepaspectratio=true,clip=true,trim=10pt 20pt 22pt 
10pt]
  {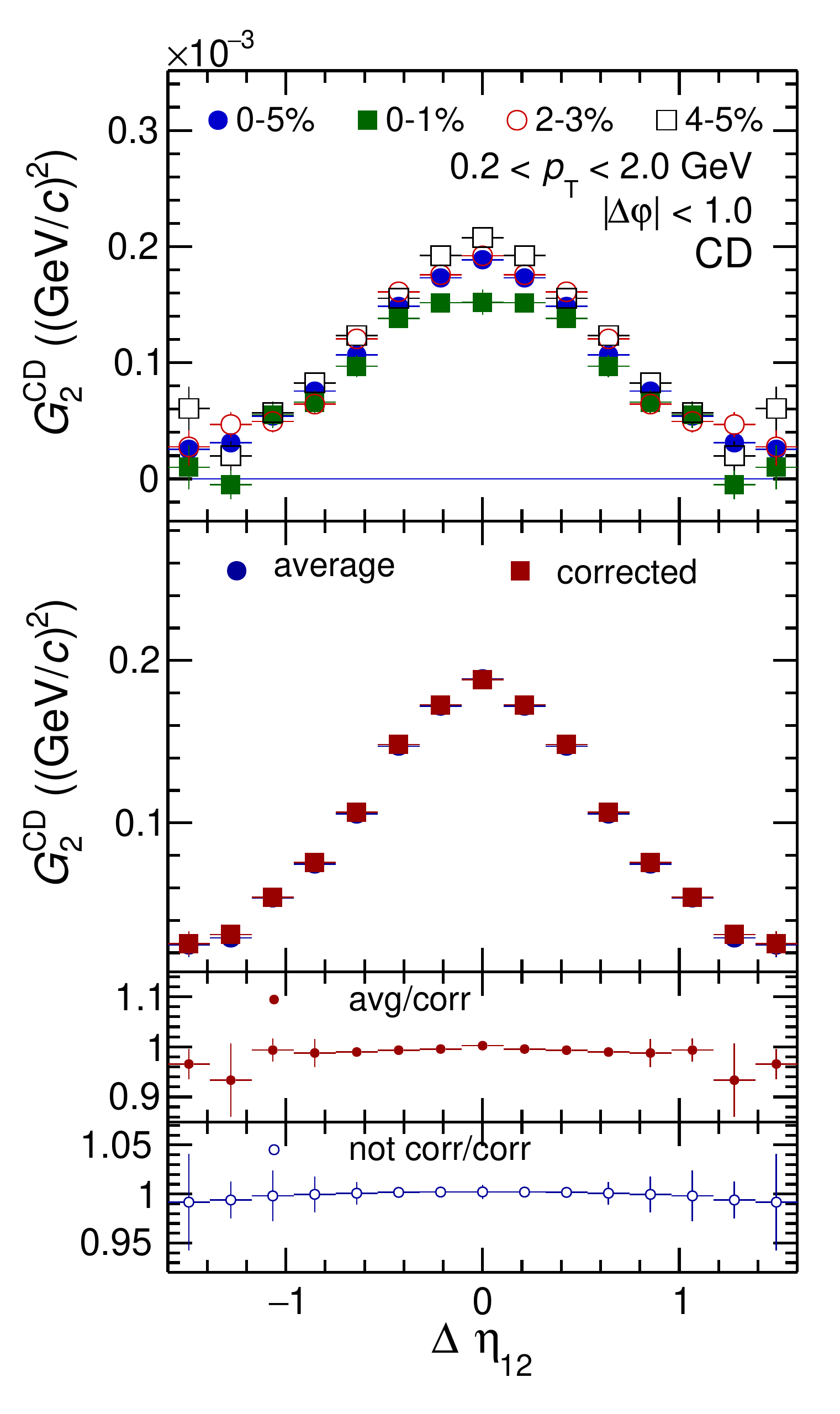} \par
\caption{\label{fig:g2etahijing}Longitudinal projection of the two-particle
transverse momentum correlations $\GtwoCI$ for default transverse momentum 
range particles (left) and for the reduced (see text) transverse momentum range 
particles (center) and $\GtwoCD$ for the default transverse momentum range
particles (right) for the 5\% most central Pb-Pb collisions produced with the
HIJING event generator. Top panels show uncorrected results for the 0--5 \% 
centrality bin together with those from the 0--1 \%, 2--3 \%, and 4--5 \% 
centrality bins, middle panels show corrected 0--5 \% centrality bin 
results compared with the weighted average of those from the 0--1 \%, 1--2 \%, 
2--3 \%, 3--4 \%, and 4-5\% centrality bins, while bottom panels show their ratio.}
\end{figure*}

\begin{figure}
 \includegraphics[scale=0.50,keepaspectratio=true,clip=true,trim=10pt 20pt 0pt 
10pt]
  {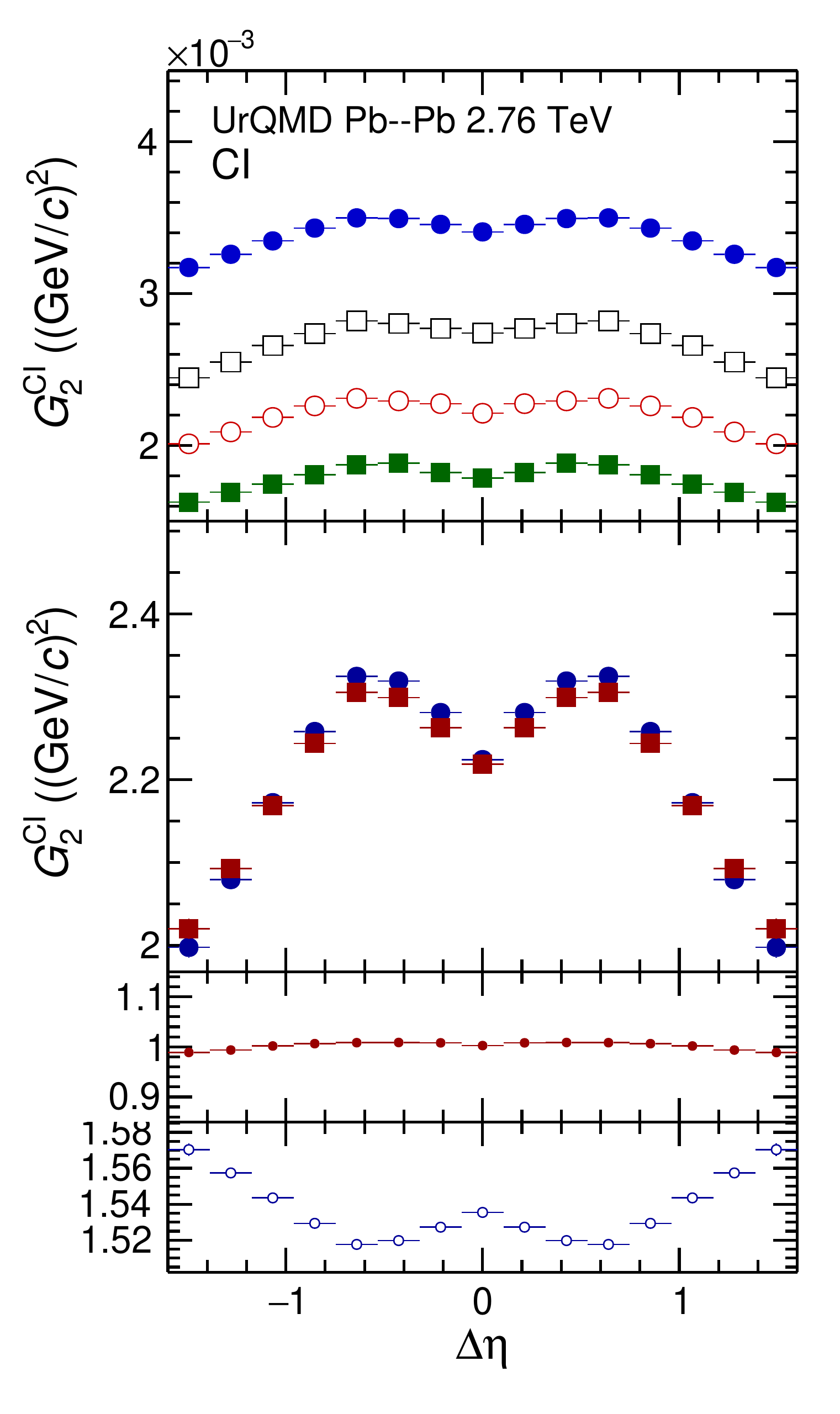}
 \includegraphics[scale=0.50,keepaspectratio=true,clip=true,trim=10pt 20pt 0pt 
10pt]
  {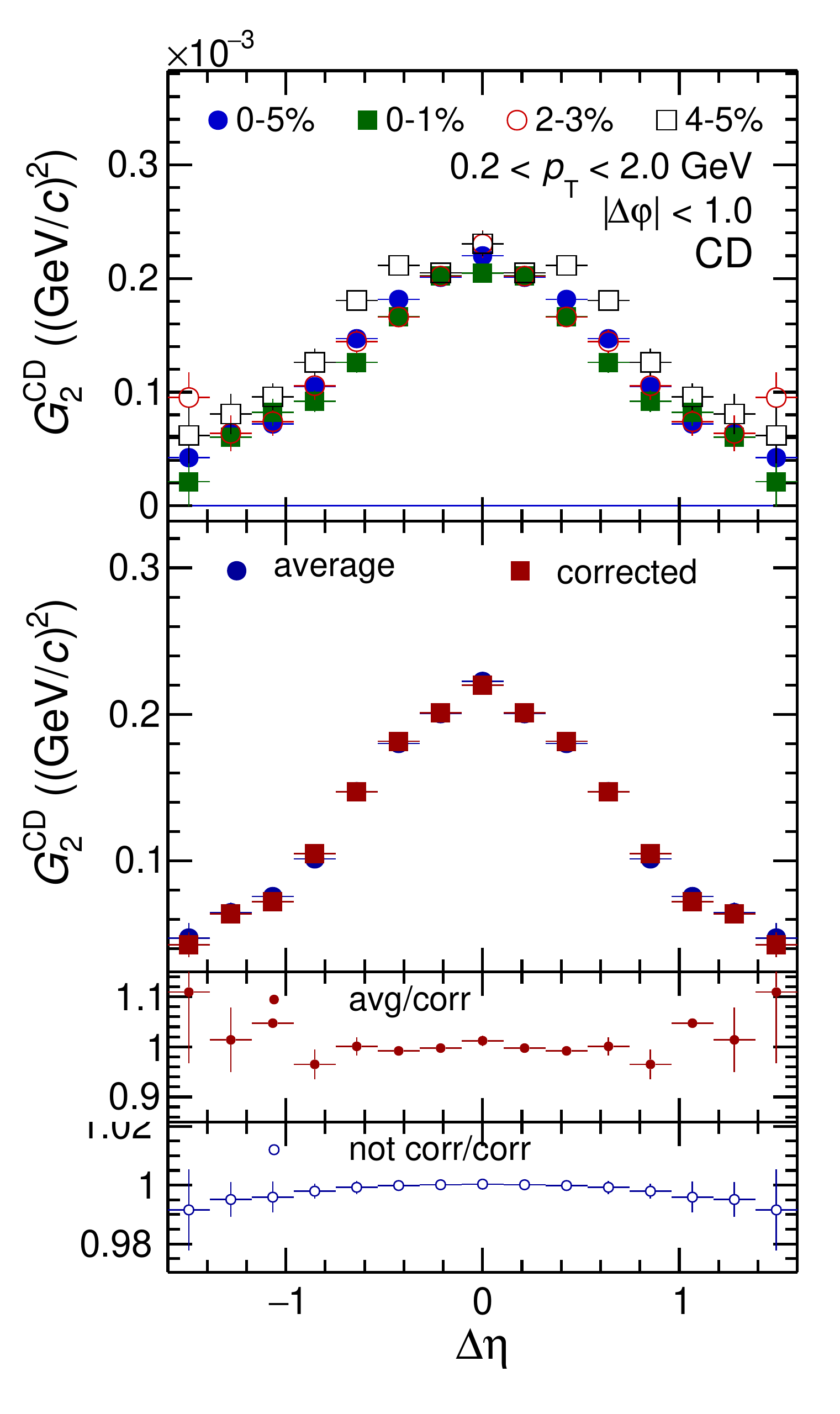} \par
\caption{\label{fig:g2etaurqmd}Longitudinal projection of the two-particle 
transverse
momentum correlations $\GtwoCI$ (left) and $\GtwoCD$ (right) for the 5\% most 
central Pb-Pb collisions produced with the UrQMD event generator. Top panels 
show uncorrected results for the 0--5 \% centrality bin together with those from 
the 0--1 \%, 2--3 \%, and 4--5 \% centrality bins, middle panels 
show corrected 0--5 \% centrality bin results compared with the weighted average 
of those from the 0--1 \%, 1--2 \%, 2--3 \%, 3--4 \%, and 4--5 \% centrality bins, while 
bottom panels show their ratio.}
\end{figure}

\begin{figure}
 \includegraphics[scale=0.50,keepaspectratio=true,clip=true,trim=10pt 20pt 0pt 
10pt]
  {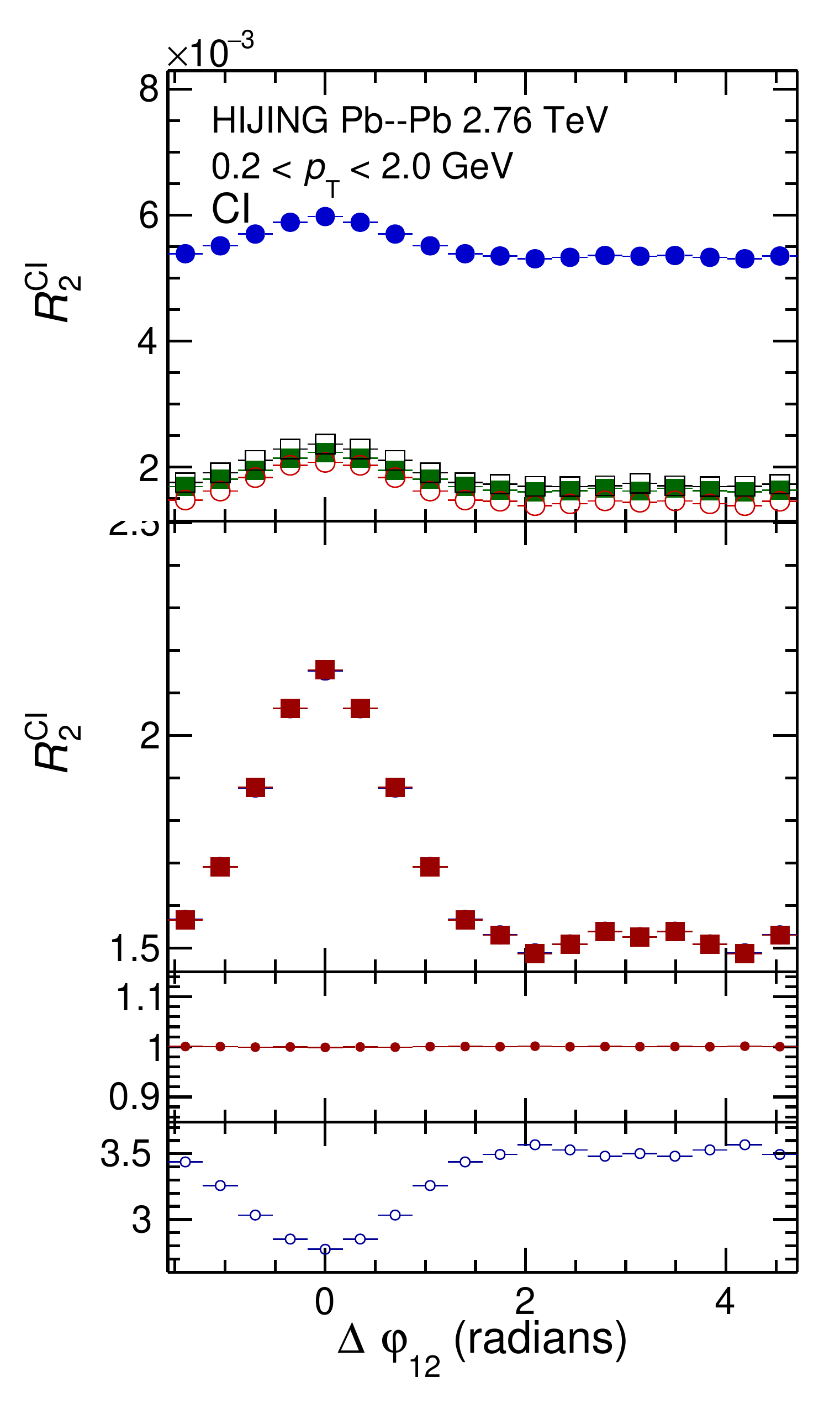}
 \includegraphics[scale=0.50,keepaspectratio=true,clip=true,trim=10pt 20pt 0pt 
10pt]
  {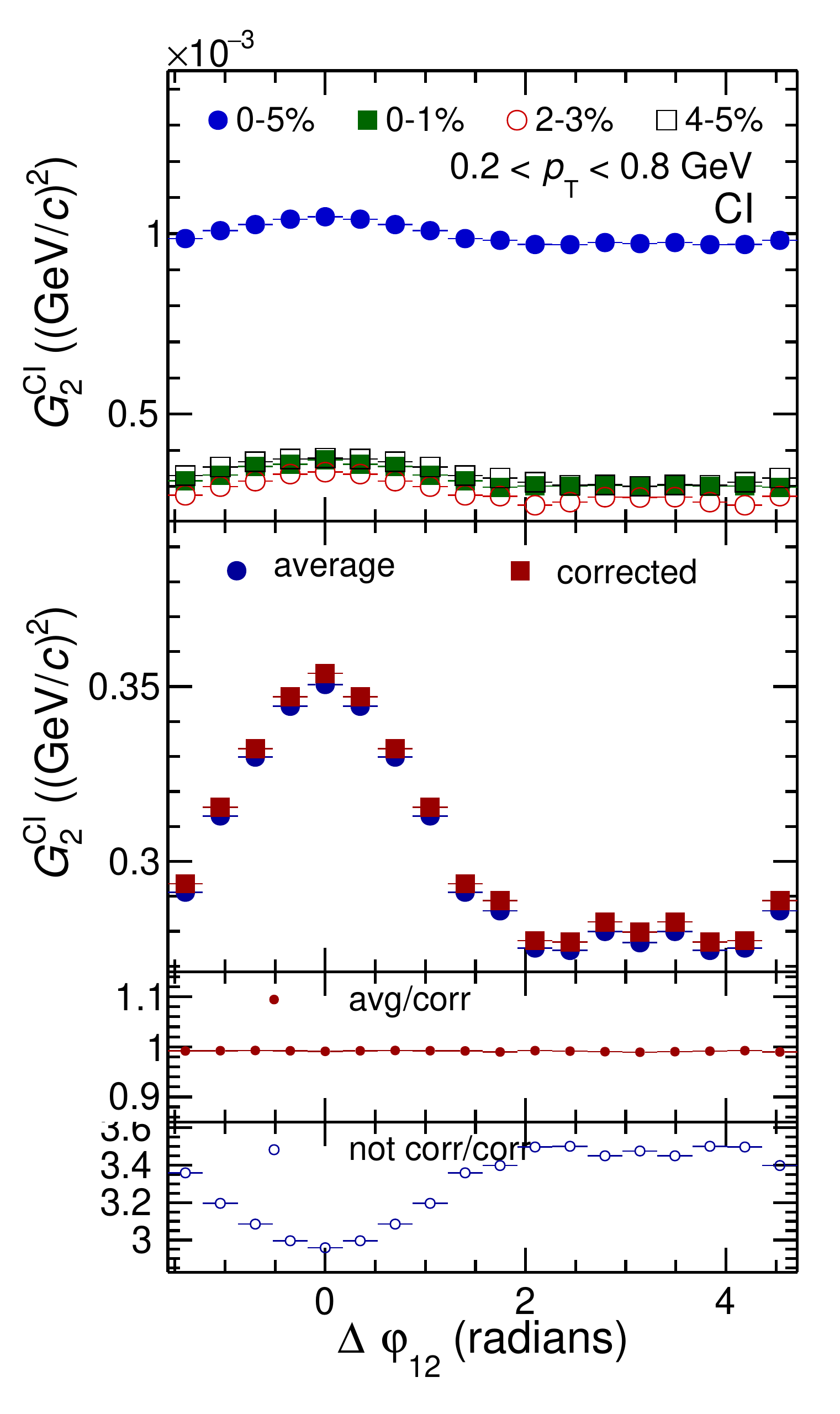} \par
\caption{\label{fig:r2g2phihijing}Azimuthal projections of the two-particle
transverse momentum correlations $\RtwoCI$ for the whole $p_{\rm T}$ range 
(left) and $\GtwoCI$ for the reduced $p_{\rm T}$ range (right) for the 5\% 
most central Pb-Pb collisions produced with the HIJING event generator. Top 
panels show uncorrected results for the 0--5 \% centrality bin together with those 
from the 0--1 \%, 2--3 \%, and 4--5 \% centrality bins, middle panels 
show corrected 0--5 \% centrality bin results compared with the weighted average 
of those from the 0--1 \%, 1--2 \%, 2--3 \%, 3--4 \%, and 4--5 \% centrality bins, while 
bottom panels show their ratio.}
\end{figure}

\begin{figure}
 \includegraphics[scale=0.50,keepaspectratio=true,clip=true,trim=10pt 20pt 0pt 
10pt]
  {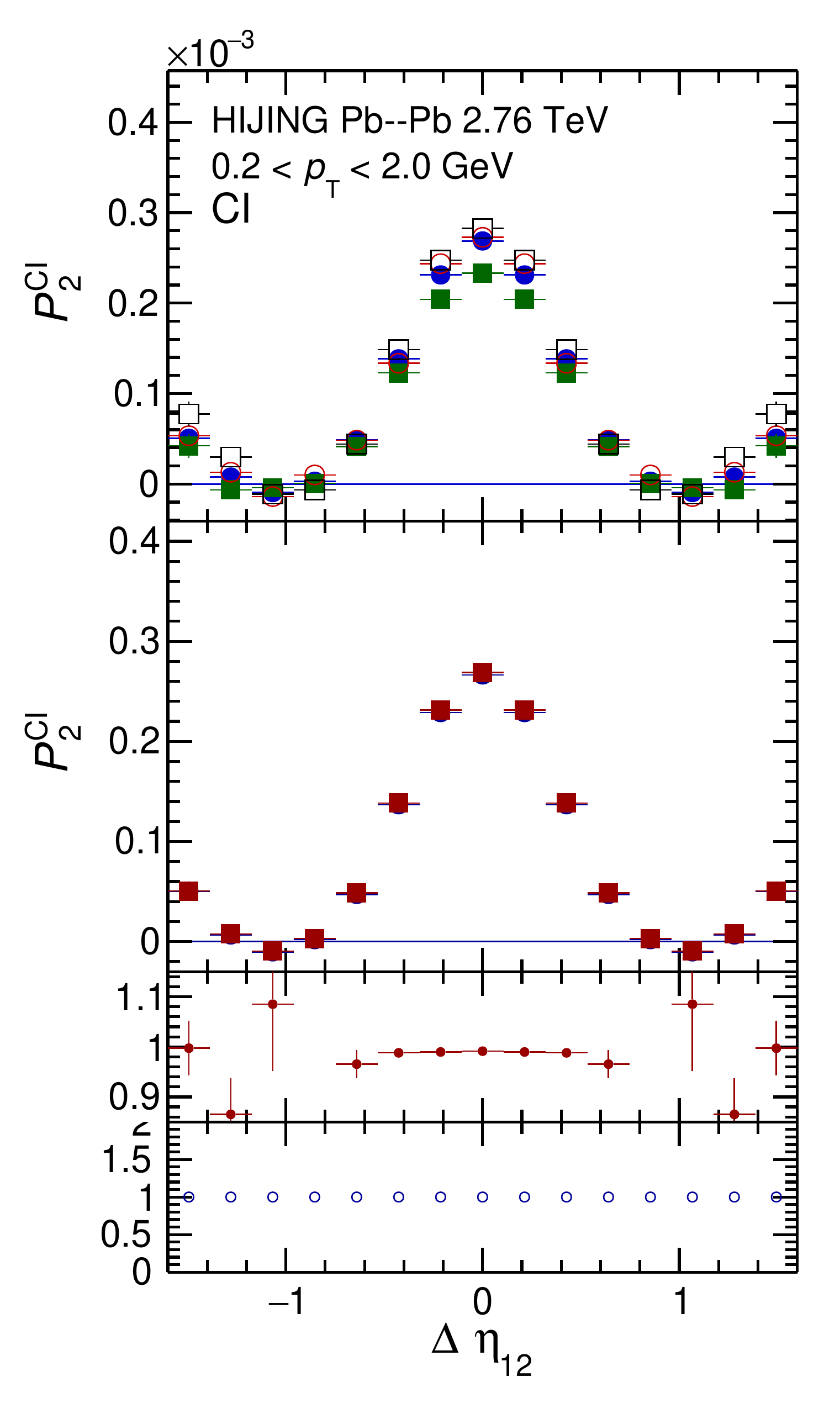}
 \includegraphics[scale=0.50,keepaspectratio=true,clip=true,trim=10pt 20pt 0pt 
10pt]
  {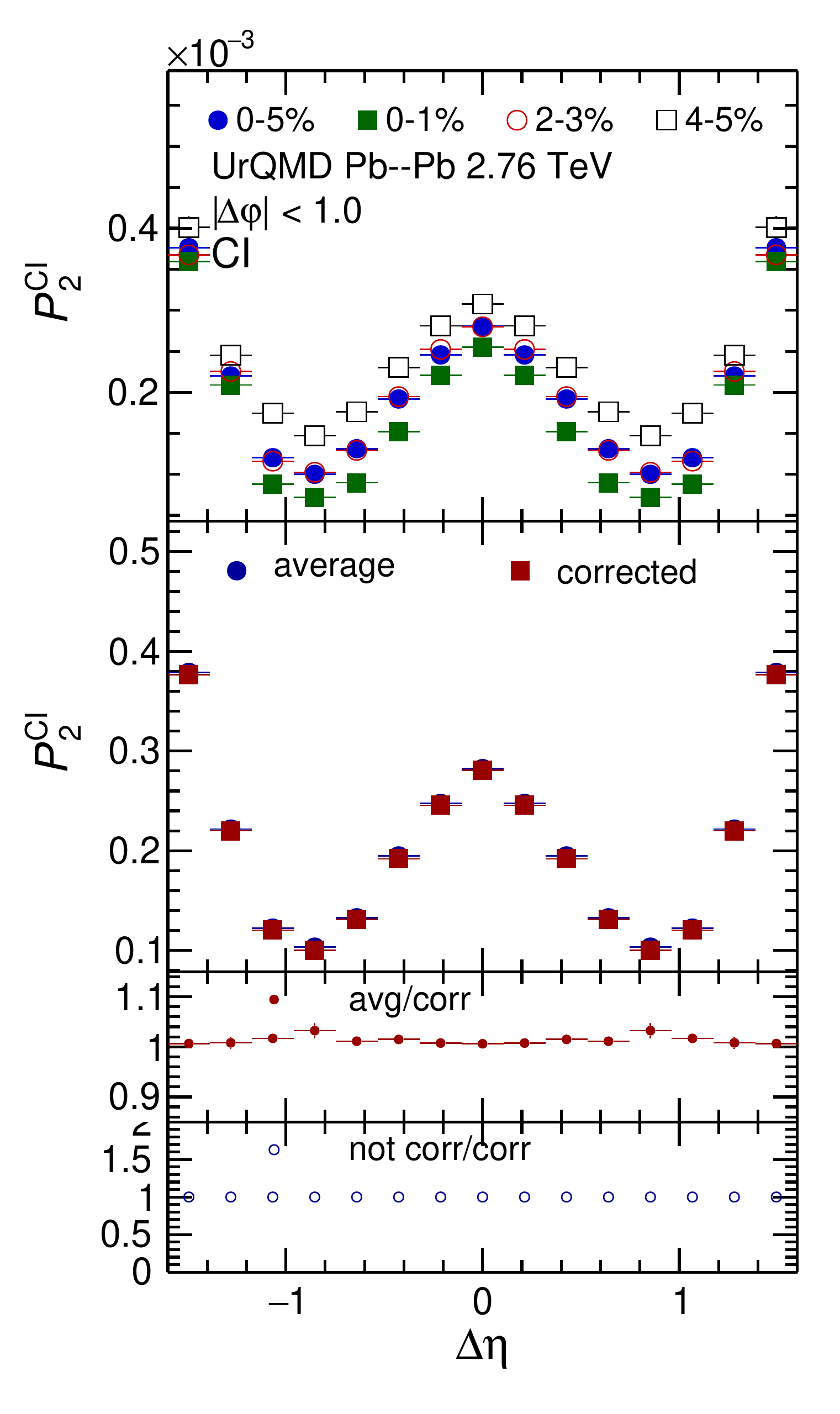} \par
\caption{\label{fig:p2etahijing}Longitudinal projection of the two-particle
transverse momentum correlation $\PtwoCI$ for the 5\% most central Pb-Pb 
collisions produced with the HIJING event generator (left) and the UrQMD event
generator (right).  Top panels show uncorrected results for the 0--5 \% 
centrality bin together with those from the 0--1 \%, 2--3 \%, and 4--5 \% 
centrality bins,   middle panels show corrected 0--5 \% centrality bin results 
compared with the weighted average of those from the 0--1 \%, 1--2 \%, 2--3 \%, 3--4 \%, 
and 4--5 \% centrality bins, while bottom panels show their ratio. 
}
\end{figure}

The accuracy of the centrality bin width correction methods defined by 
 Eqs.~(\ref{eq:r2corr}) and (\ref{eq:g2corr}) is tested using HIJING and UrQMD simulations 
of 5\% most central  central Pb-Pb collisions at $\snn = 2.76~\text{TeV}$.
Overall,  $6.5\times10^4$ and $10^5$ Pb-Pb collisions from HIJING and UrQMD were used 
for this study. 
The collision centrality classification used for  HIJING events mimics  the 
method used by the  ALICE collaboration~\cite{Abelev:2013qoq,Abelev:2014ffa} and is based on the 
number of charged particles in the pseudo-rapidity ranges $2.8 < \eta < 5.1$ and 
$-3.7 < \eta < -1.7$ corresponding to the acceptance of the ALICE V0 detectors.
For UrQMD simulations, the centrality selection is based on sharp cuts in the 
impact parameter of the collision, as given in~\cite{Abelev:2013qoq}.
The impact of the centrality estimation method is not  considered in this work. It is assumed that the centrality is properly extracted and that the method employed to estimate it does not influence the results. 
In actual measurements, the  impact of the centrality estimation method
shall be inferred with the usual methods for systematic uncertainties extraction.
The $R_2$, $G_2$, and $P_2$ correlators for charged particle combinations ($+-$, $-+$, $--$ and $++$) are first determined as four-dimensional functions based on generator level charged hadrons in the ranges $0.2 \le p_{\rm T} \le 2.0$ GeV/$c$ and $|\eta| < 0.8$ using Eqs.~(\ref{eq:r2corr}), (\ref{eq:g2corr}), and (\ref{eq:p2corr}). 
From these, the charge independent (CI) and charge dependent (CD) particle combinations are calculated according to
\be
C^{\rm CI} &=& \frac{1}{4}\left[ C^{\left(+-\right)} + C^{\left(-+\right)} 
    + C^{\left(--\right)} + C^{\left(++\right)} \right] \\
C^{\rm CD} &=& \frac{1}{4}\left[ C^{\left(+-\right)} + C^{\left(-+\right)} 
    - C^{\left(--\right)} - C^{\left(++\right)} \right]\text{.}
\ee
where $C$ stands for any of the $R_2$, $G_2$, and $P_2$ correlators. Projections 
of these correlators onto  $\Delta\eta$ and $\Delta\varphi$ were computed
using the integer arithmetic technique described in~\cite{Ravan:2013lwa}.  

Figure~\ref{fig:r2etahijing} presents $\Delta \eta$ projections of the 
$\RtwoCI$ (left) and $\RtwoCD$ (right) correlation functions based on 5\% 
most central  HIJING events. 
The top panels of the figure present uncorrected results for 0--5 \% (blue), 0--1 \%
(green), 2--3 \% (red), and 4--5 \% (black). The bottom panels displays 0--5 \% 
centrality results corrected with Eq.~(\ref{eq:r2corr}) (blue) compared to the 
weighted mean of the correlators obtained for 0--1 \%, 1--2 \%, 2--3 \%, 3--4 \%, and 
4--5 \% centralities (red) and their ratio.
One finds that the results corrected with Eq.~(\ref{eq:r2corr}) agree with those
obtained with the weighted mean within 1\% for both $\RtwoCI$  and 
$\RtwoCD$. 

Similar results are presented in Fig.~\ref{fig:r2etaurqmd} for the UrQMD 
model. In this case, agreement between the 0--5 \% corrected $R_2$ 
correlators and the weighted mean of the $R_2$ correlators obtained in 0--1 \%, 
1--2 \%, 2--3 \%, 3--4 \%, and 4--5 \% centralities is found to be also within 1\% for both 
$\RtwoCI$  and $\RtwoCD$. 
Thus, based on the HIJING and UrQMD model simulations presented, it is concluded
that Eq.~(\ref{eq:r2corr}) enables reasonably accurate corrections of the $R_2$ 
correlators in the context of these two models. Given these models provide
relatively realistic representations of single and pair particle spectra, the correction method embodied in Eq.~(\ref{eq:r2corr}) should provide 
reasonably reliable bin-width corrections of $R_2$ correlation functions 
measured at any heavy ion collider.
  
Figures~\ref{fig:g2etahijing} and~\ref{fig:g2etaurqmd} present $\Delta \eta$ 
projections of the $\GtwoCI$  and $\GtwoCD$ correlation functions based 
on  5\% most central events from HIJING and UrQMD models respectively. 
Similarly as in previous figures, the top panels display uncorrected $G_2$ 
correlators for 0--5 \% (blue), 0--1 \% (green), 2--3 \% (red), and 4--5 \% (black) 
while the bottom panels display 0--5 \% centrality results corrected with 
Eq.~(\ref{eq:g2corr}) (blue) compared to the weighted mean of the $G_2$ 
correlators obtained for 0--1 \%, 1--2 \%, 2--3 \%, 3--4 \%, and 4--5 \% centralities 
(red) and their ratio. 
Comparing the corrected results with the weighted averages for HIJING events 
one finds them within 7\% for $\GtwoCI$ and within 1\% for $\GtwoCD$ 
while the comparison for UrQMD events leaves them within 1\% for both 
$\GtwoCI$ and $\GtwoCD$.

The discrepancy between the correction level achieved for $\GtwoCI$ for 
collisions simulated with HIJING compared with those  produced using UrQMD is  
hardly satisfactory, especially considering the 1\% correction precision  
achieved, with both models, for the  $\RtwoCI$ correlator. 
Given the transverse momentum enters explicitly into the expression of  the 
$G_2$ correlator, but not in $R_2$, to determine whether the $p_{\rm 
T}$ range considered may influence the correction precision achievable with 
Eq.~(\ref{eq:g2corr}) the above study is repeated for charged particles 
in the range 0.2--0.8~\gevc.
Central panels of Fig.~\ref{fig:g2etahijing} display $\Delta \eta$ 
projections of the $\GtwoCI$  correlation function based on 5\% most central 
events from HIJING model for particle within the transverse momentum range 
0.2--0.8~\gevc. As before, the top panel shows a  comparison of the 
correlators calculated within  the 0--5 \% centrality bin with  those from the
0--1 \%, 2--3 \%, and 4--5 \% centrality bins, while the bottom panel displays 0--5 \% 
centrality results corrected with Eq.~(\ref{eq:g2corr}) compared to the 
weighted mean of the values obtained in the 0--1 \%, 1--2 \%, 2--3 \%, 3--4 \%, and 
4--5 \% collision centrality bins. It is found that  the 0--5 \% centrality corrected 
result is now within 1\% of the weighted mean of those obtained with  finer 
centrality bins. The centrality dependence of the single- and two-particle probability densities [and their  $p_{\rm T}$ weighted counterparts, Eqs.~(\ref{eq:P1pT}) and
(\ref{eq:P2pTpT})] is then examined, and found that those obtained within the  0.2--0.8~\gevc\ range
feature a quantitatively smaller dependence of collision centrality than those
integrated in the 0.2--2.0~\gevc\ range. The former range consequently better satisfies
the condition, implicit in Eq.~(\ref{eq:G2Factor}), that the densities be approximately independent of 
collision centrality, within each centrality bin $k$. That 
these densities, calculated within the range 0.2--2.0~\gevc\ with UrQMD events, similarly
exhibit very little shape dependence on collision centrality is additionally verified. 
This analysis confirms the relevance of the proper behavior of this magnitudes established in Sec.~\ref{sec:G2Correlators}.
As conclusion then, the finite centrality bin width correction, included in 
Eq.~(\ref{eq:g2corr}), under the criteria established for its deduction, 
provides a reasonably accurate correction technique for $G_2$ correlators to 
account for collision centrality average when using finite collision centrality 
bins. 
  
It is worth highlighting that, consistently with what was naively expected, the centrality bin width effect is completely charge independent. Indeed, the amplitude of both
$\RtwoCI$ and $\GtwoCI$ are substantially affected by the wide centrality bin averaging while $\RtwoCD$ and $\GtwoCD$ only experience a minor shift owing (1) to the very small differences 
between the  $\RtwoLS$ and  $\RtwoUS$ correlator dependence on particle densities and on the transverse momentum of the particles, and (2) to the approximately equal number of US and LS pairs observed at large 
$\snn$ in the central rapidity range. 

For the sake of completeness, the behavior of  the correction procedures on the 
$\Delta \varphi$ projections of $\RtwoCI$ (for the whole $p_{\rm T}$ range) 
and $\GtwoCI$ (for the reduced $p_{\rm T}$ range) correlators are shown in 
Fig.~\ref{fig:r2g2phihijing}. The correction procedures  yield 
essentially perfect corrections of these $\Delta \varphi$ projections for both 
correlators. 
Additionally, the applicability of Eq.~(\ref{eq:p2corr}) for $\PtwoCI$ 
correlators is demonstrated in Fig.~\ref{fig:p2etahijing} where it is shown that 
the $\PtwoCI$ correlator determined with a 0--5 \% centrality  bin is 
virtually identical to that obtained with a weighted average of the 
correlators calculated from the 0--1 \%, 1--2 \%, 2--3 \%, 3--4 \%, and 4--5 \% collision
centrality bins.

Finally, in order to obtain  a quantitative assessment of the precision of the 
centrality bin width correction procedures embodied by 
Eqs.~(\ref{eq:r2corr}), (\ref{eq:p2corr}), (\ref{eq:g2corr}), the ratios of 
the centrality bin width corrected correlators to those obtained with weighted 
averages of correlators obtained in fine width centrality bins, shown in 
Figs.~1--6, are fitted with a constant polynomial (POL0 in ROOT~\cite{Brun:1997pa}). 
Fitted values of the ratios, corresponding to  ratios averaged across the 
$\Delta\eta$ and $\Delta\varphi$ fiducial ranges,  are listed in  
Table~\ref{tab:pol0val}.
\begin{table*}
\caption{\label{tab:pol0val} POL0 fit to the ratio 0--5 \% 
centrality corrected versus 
0--1 \%, 1--2 \%, 2--3 \%, 3--4 \%, and 4--5 \% centralities weighted mean.}
\begin{ruledtabular}
\begin{tabular}{lccccccc}
 Model & proj. & $\RtwoCI$ & $\RtwoCD$ & $\PtwoCI$ & $\PtwoCD$ 
  & $\GtwoCI$ & $\GtwoCD$ \\
\hline
HIJING & $\Delta \eta$      & $1.000 \pm 0.001$ & $0.995 \pm 0.003$
  &  $0.989 \pm 0.002$ &  $0.999 \pm 0.004$ 
  &  $0.931 \pm 0.001$ &  $0.997 \pm 0.002$ \\
HIJING & $\Delta \varphi$   & $1.000 \pm 0.001$ & $0.988 \pm 0.003$
  &  $0.987 \pm 0.003$ &  $0.994 \pm 0.008$ 
  &  $0.935 \pm 0.001$ &  $0.985 \pm 0.004$ \\
UrQMD & $\Delta \eta$      & $1.001 \pm 0.0003$ & $1.005 \pm 0.0016$
  & $1.010 \pm 0.0012$ &  $0.893 \pm 0.0435$ 
  & $1.007 \pm 0.001$  & $1.002 \pm 0.002$ \\
UrQMD & $\Delta \varphi$   & $0.990 \pm 0.0002$ & $1.047 \pm 0.0027$
  & $1.022 \pm 0.0007$ & $1.040 \pm 0.0222$ 
  & $1.008 \pm 0.001$  & $0.985 \pm 0.003$ \\
\multicolumn{1}{l}{HIJING low $p_{\rm T}$} & $\Delta \eta$ & & & &
  &  $0.995 \pm 0.001$ &  $0.993 \pm 0.006$ \\
\multicolumn{1}{l}{HIJING low $p_{\rm T}$} & $\Delta \varphi$ & & & &
  &  $0.991 \pm 0.001$ &  $0.989 \pm 0.002$ \\
\end{tabular}
\end{ruledtabular}
\end{table*}

Overall, it is found that  the averaged ratios have values 
consistent with unity within statistical errors (i.e., within a $\pm 2\sigma$, 
96\% confidence interval), thereby implying that within the statistical 
precision achieved in this work, it can be concluded that the correction procedures 
do not introduce a significant bias to the correlation functions. 
Notable exceptions are the values obtained with both $\Delta\eta$ and 
$\Delta\phi$ projections for $\GtwoCI$ when this correlator is calculated for 
charged particles in the range $0.2\le p_{\rm T}\le 2.0$ \gevc\ with HIJING. However, the ratios of these projections are consistent with unity when the 
correlator is calculated in a narrower $p_{\rm T}$ range in line with the 
demanded behavior of the probability density distributions.
A large discrepancy is also seemingly observed for the $\PtwoCD$ correlator
obtained with the UrQMD simulation. However, the overall magnitude of $\PtwoCD$ 
predicted by  UrQMD  is a seventh of that obtained with HIJING events and correlation amplitudes
are nearly vanishing, within statistical accuracy, across a wide $\Delta\eta_{12}$ range. 
A proper evaluation of the correction is thus more challenging in this case. 

\section{Summary}
\label{sec:Summary}

Studies of the centrality or particle multiplicity evolution of the integral and amplitudes of two-particle number and transverse momentum differential correlations, as tools to gain insights into particle production and transport dynamics in heavy-ion collisions, are a growing research field of interest, as recent analyses from the ALICE collaboration show~\cite{Adam:2017ucq,Acharya:2018ddg}.
The bias introduced on the amplitudes and integrals of $R_2$, $G_2$, and $P_2$ correlators when measured directly in wide collision centrality bins, was shown.
Correction techniques to be applied to those correlators, when the need for them to be measured on such centrality bins is in place, were described. 
These correction techniques were tested with HIJING and UrQMD simulations of Pb-Pb collisions at $\snn = 2.76~\text{TeV}$. They  enable a precision of the order of 1\%, or better, when using 5\% wide centrality bins. Clearly, the reached precision has a direct impact on both the amplitude and the integral of the correlation functions which will allow a better description or constrain of the underlying physics.

\newenvironment{acknowledgement}{\relax}{\relax}
\begin{acknowledgement}
\section*{Acknowledgements}
The authors thank the ALICE collaboration for access to a sample of simulated HIJING events used in this analysis and the GSI Helmholtzzentrum f\"ur Schwerionenforschung for providing the computational resources needed for producing the UrQMD events used in this analysis. This work was supported in part by the United States Department  of Energy, Office of Nuclear Physics (DOE NP), United States of America under Award No. DE-FG02-92ER-40713 and  by Consejo Nacional de Ciencia y Tecnolog\'{\i}a (CONACYT), through Fondo de Cooperaci\'{o}n Internacional en Ciencia y Tecnolog\'{\i}a (FONCICYT) and Direcci\'{o}n General de Asuntos
del Personal Acad\'{e}mico (DGAPA), Universidad Nacional Aut\'{o}noma de M\'{e}xico, Mexico.
\end{acknowledgement}

\bibliography{binWidthCorr}

\end{document}

%% file: defs.tex
\newcommand{ \be }{\begin{linenomath*}\begin{eqnarray}}
\newcommand{ \ee }{\end{eqnarray}\end{linenomath*}}
\newcommand{ \la }{\langle}

\newcommand{ \ra }{\rangle}

\def \mean#1 {{\la #1 \ra}}

\newcommand{ \gevc }{GeV/\textit{c}}

\newcommand{\RtwoLS}{R_2^{\rm LS}}
\newcommand{\RtwoUS}{R_2^{\rm US}}
\newcommand{\RtwoCI}{R_2^{\rm CI}}
\newcommand{\RtwoCD}{R_2^{\rm CD}}

\newcommand{\PtwoCI}{P_2^{\rm CI}}
\newcommand{\PtwoCD}{P_2^{\rm CD}}

\newcommand{\GtwoCI}{G_2^{\rm CI}}
\newcommand{\GtwoCD}{G_2^{\rm CD}}

\newcommand{\PDF}{{\mathbb{P}}}

\definecolor{dgreen}{cmyk}{1.,0.,1.,0.2}        
\definecolor{orange}{cmyk}{0.,0.353,1.,0.}    

\def \snn {\sqrt{\textit{s}_{_{\rm NN}}}}

\long\def\/*#1*/{}